\documentclass[times,final]{elsarticle}

\usepackage{jcomp}
\usepackage{framed,multirow}

\usepackage{amssymb}
\usepackage{latexsym}

\usepackage{kotex}
\usepackage{cleveref}
\usepackage{multirow}
\usepackage{booktabs}

\usepackage{url}
\usepackage{xcolor}
\definecolor{newcolor}{rgb}{.8,.349,.1}

\journal{Journal of Computational Physics}

\begin{document}

\verso{Kim and Lee}

\begin{frontmatter}

\title{Deep unsupervised learning of turbulence for inflow generation at various Reynolds numbers}%

\author[1]{Junhyuk Kim}

\author[1,2]{Changhoon Lee\corref{cor1}}
\ead{clee@yonsei.ac.kr}
\cortext[cor1]{Corresponding author.}

\address[1]{Department of Mechanical Engineering, Yonsei University, Seoul 03722, Korea}
\address[2]{Department of Computational Science and Engineering, Yonsei University, Seoul 03722, Korea}

\received{}
\finalform{}
\accepted{}
\availableonline{}
\communicated{}

\begin{abstract}
A realistic inflow boundary condition is essential for successful simulation of the developing turbulent boundary layer or channel flows. Recent advances in artificial intelligence (AI) have enabled the development of an inflow generator that performs better than the synthetic methods based on intuitions. In the present work, we applied generative adversarial networks (GANs), a representative of unsupervised learning, to generate an inlet boundary condition of turbulent channel flow. Upon learning the two-dimensional spatial structure of turbulence using data obtained from direct numerical simulation (DNS) of turbulent channel flow, the GAN could generate instantaneous flow fields that are statistically similar to those of DNS. Surprisingly, the GAN could produce fields at various Reynolds numbers without any additional simulation based on the trained data of only  three Reynolds numbers. This indicates that the GAN could learn the universal nature of Reynolds number effect and might reflect other simulation conditions. Eventually, through a combination of the GAN and a recurrent neural network (RNN), we developed a novel model (RNN-GAN) that could generate time-varying fully developed flow for a long time. The spatiotemporal correlations of the generated flow are in good agreement with those of the DNS. This proves the usefulness of unsupervised learning in the generation of synthetic turbulence fields. 
\end{abstract}


\end{frontmatter}

\section{Introduction}\label{sec:intro}

The recent development of artificial intelligence (AI), especially deep learning \cite{LeCun2015}, has inspired many researchers in a wide range of disciplines, and various applications of deep learning have been carried out in their quest for understanding complex processes. Turbulence, which is one of the complex phenomena that are strongly nonlinear and unpredictable, might be a good target for deep learning. In particular, one of the unsupervised learning algorithms known as generative adversarial network (GAN) proposed by \citet{Goodfellow2014} could be used to generate a field statistically similar to turbulence. In this study, we developed a generative network adopting GAN to provide an inflow turbulence field for the simulation of the developing channel turbulence, which is statistically similar to the field obtained from the direct numerical simulation (DNS) of a fully developed turbulent channel flow.

A proper inflow boundary condition, which is essential for simulations of developing flows such as channel turbulence or boundary layer flow, needs to be generated. The commonly used generation methods can be divided into two: recycling inflow generation and synthetic inflow generation, as reviewed by \citet{Tabor2010} and \citet{Wu2017}. The recycling method \cite{Kim1987,Spalart1988,Lund1998} involves carrying out an auxiliary simulation under a periodic boundary condition and using the fields in a cross-sectional plane to the mean flow direction as the inlet boundary condition of the main simulation. This method has an advantage that it can produce a flow field with accurate spatial and temporal correlations under the assumption that a sufficiently long domain size is used in the auxiliary simulation. However, it requires a large computational cost. A typical example of the auxiliary simulation is a DNS of turbulent channel flow with the periodic boundary condition. On the other hand, the synthetic inflow turbulence generation method involves generating turbulence quickly under a certain constraint to provide realistic turbulence fields in a short time. Several models, such as synthetic random Fourier method \cite{Le1997}, synthetic digital filtering method \cite{Klein2003}, and synthetic coherent eddy method \cite{Jarrin2006,Oh2019}, have been proposed to generate flow with a proper spatiotemporal correlation to prevent rapid dissipation of turbulence \cite{Keating2004}. Although these approaches produce reasonable inflow fields, the generation methods depend on ad hoc assumptions. For example, the size of the synthetic eddy should be carefully selected by humans based on their intuitions and experiences, otherwise a very long transition section might be needed \cite{Jarrin2006}. Therefore, we need an efficient generation algorithm that can quickly produce turbulence fields with DNS-quality statistics. 

Recently, it was reported that the supervised learning of turbulence, which predicts the next step information based on the previous step information, is feasible. This method can be utilized for the inflow turbulence generation problem. In general, convolution neural networks (CNN) are used to better represent the spatial correlation of flow \cite{King2018,Mohan2019,Fukami2019,Lee2017,Lee2018,Ruettgers2019}. However, the typical CNN structure \cite{Lee2017} generates blurred fields over time when the recursive prediction, which uses the predicted information as the input, is carried out. To solve this problem, an autoencoder (AE) architecture composed of an encoder and a decoder was used \cite{King2018,Mohan2019,Fukami2019}. Owing to the feature extraction during encoding, the AE becomes less sensitive to the input; hence, the output is almost statistically stationary. \citet{Fukami2019} showed that the AE becomes applicable to the turbulence generation problem after training the DNS data of channel flow at $Re_\tau=180$. Although deep learning for inflow generation was successfully carried out, there are still issues to be resolved. The turbulence generated by the AE is not well represented because the output is not uniquely determined by the input, and the AE architecture commonly produces a blurred field. In addition, there is a crucial disadvantage that a fully developed field is required as an initial input.

In this study, in order to resolve the above issues, we applied an unsupervised learning to the inflow generation. In unsupervised learning, the generated data can be statistically the same as the training data. Also, another great advantage is that the initial input field is not required. In particular, GANs, which were proposed by \citet{Goodfellow2014} as one of the unsupervised learning methods, have been improved steadily \cite{Radford2015,Arjovsky2017,Gulrajani2017,Karras2017,Roth2017,Mescheder2018,Karras2018}. Although GANs have been used a little in fluid dynamics, most of them were supervised learning problems \cite{Lee2017,Ruettgers2019,Xie2018}. Unsupervised learning using GAN has been rarely studied (except \cite{Wu2019}), and it has considerable potential in the development of turbulence generation. In the present work, we trained the GAN to generate instantaneous flow fields using training data collected by the channel flow DNSs for three Reynolds numbers. Then, we extended the trained GAN to an inflow generator by combining it with a recurrent neural network (RNN) to generate the time-variation of flow. This RNN-GAN is capable of generating a time-varying fully developed flow at various Reynolds numbers for a long time. Through this study, we show that deep unsupervised learning would be a useful tool for the development of turbulence generation researches.

This article consists of a simulation part (Section \ref{sec:simulation}), a deep learning part (Section \ref{sec:unsup}), and conclusion (Section \ref{sec:conclusions}). In Section \ref{sec:simulation}, we carry out DNSs for three different Reynolds numbers to collect the training data for the generation algorithm.  In Section \ref{sec:unsup}, the details of our deep learning methods and results are presented. The generation of instantaneous flow fields at various Reynolds numbers using the GAN and the generation of time-varying flow using RNN-GAN are described in Sections \ref{subsec:GAN} and \ref{subsec:RNN-GAN}, respectively.

\section{Simulations for data collection}\label{sec:simulation}
The basic idea is to construct an unsupervised deep learning network to generate fields that possess statistics similar to those of DNS fields. Then, the generated fields can be used as the inlet boundary condition for developing flow simulations. In order to collect data for unsupervised learning, DNSs of a fully developed turbulent channel flow with the temperature field were carried out.  In the channel flow, the mean pressure gradient drives the mean flow in the $x$-direction, and the temperature difference between two walls induces a heat flux in the wall-normal direction. The governing equations are the continuity equation, the Navier--Stokes equations, and the energy equation as follows:
\begin{equation}
{\partial u_i \over \partial x_i} =0,
\end{equation}
\begin{equation}
{\partial {u_i} \over \partial t} + u_j {\partial u_i  \over \partial x_j} = -{\partial p \over \partial x_i} + {1 \over Re_\tau}{\partial^2{u_i} \over \partial x_j \partial x_j},
\end{equation}
\begin{equation}
{\partial {T}\over{\partial t}} + u_j {\partial T \over \partial x_j }  = {1\over Pr~ Re_\tau}{\partial^2{T} \over \partial x_j \partial x_j},
\end{equation}
where $x_i$, $u_i$, and $T$ are dimensionless variables of the channel half width ($\delta$), friction velocity ($u_\tau$), and the half temperature difference ($\Delta T$) between the two walls, respectively. Here, $x_1$ ($x$), $x_2$ ($y$), and $x_3$ ($z$) denote the streamwise, wall normal, and spanwise directions, and the corresponding velocity components are $u_1$ ($u$), $u_2$ ($v$), and $u_3 $($w$), respectively. The non-dimensional input parameters of simulation are the friction Reynolds number ($Re_\tau$), which is defined by $u_\tau$, $\delta$, and kinetic viscosity, and the molecular Prandtl number ($Pr$), which is defined by the ratio between the kinetic viscosity and thermal diffusivity.
\begin{table}
\caption{Simulation parameters. $L$ and $N$ are the domain size and the number of grid points, respectively. $\Delta y^+_w$ and $\Delta y^+_c$ are the resolutions near the wall and at the center of the channel, respectively.}\label{tab:1}
 \begin{center}
  \begin{tabular}{cccccccccc}
   \hline
    $Re_\tau$ & $Pr$ &  $L_x \times L_y \times L_z$  &  $L_y^+ \times L_z^+$ & $N_x \times N_y \times N_z$ & $\Delta x^+$ & $\Delta z^+$ & $\Delta y^+_w$ & $\Delta y^+_c$ & $\Delta t^+_{DNS}$ \\ \hline 
    $180$   & $0.71$ & $4\pi\delta\times2\delta\times2\pi\delta$ & $360.0\times 1130.97$ & $192\times129\times192$ & 11.78 & 5.89 & 0.054 & 4.42 & 0.090 \\
    $360$   & $0.71$ & $4\pi\delta\times2\delta\times\pi\delta$ & $720.0\times 1130.97$ & $384\times193\times192$ & 11.78 & 5.89 & 0.048 & 5.89 & 0.045\\
    $540$   & $0.71$ & $4\pi\delta\times2\delta\times{2\over 3}\pi\delta$ & $1080.0\times 1130.97$ & $576\times257\times192$ & 11.78 & 5.89 & 0.041 & 6.63 & 0.045\\
    \hline
   \end{tabular} 
 \end{center}
\end{table}

Periodic boundary conditions are applied in the streamwise and spanwise directions, and the no-slip and constant temperature boundary conditions are imposed on the wall. A pseudo-spectral method with Fourier expansion in the horizontal directions and Chebyshev-tau method in the wall-normal direction is used. The third-order Runge--Kutta scheme and Crank--Nicolson scheme are used for the time advancement of the nonlinear terms and viscous terms, respectively. We carried out the simulations at three Reynolds numbers, and the domain size and the number of grid points for each Reynolds number are listed in Table \ref{tab:1}; the resolutions in wall units are similar to those of \citet{Moser1999}. The domain size in the spanwise direction at each Reynolds number is maintained the same in wall units. On the other hand, the domain size in the  streamwise direction increases with Reynolds number in order to prevent the problem that the time-correlation becomes too high due to the periodic boundary condition, and to collect more diverse data. The simulation results were validated with those of \citet{Kim1987}, \citet{Garcia2011}.

\begin{figure}
  \centerline{
  \includegraphics[width=0.95\columnwidth]{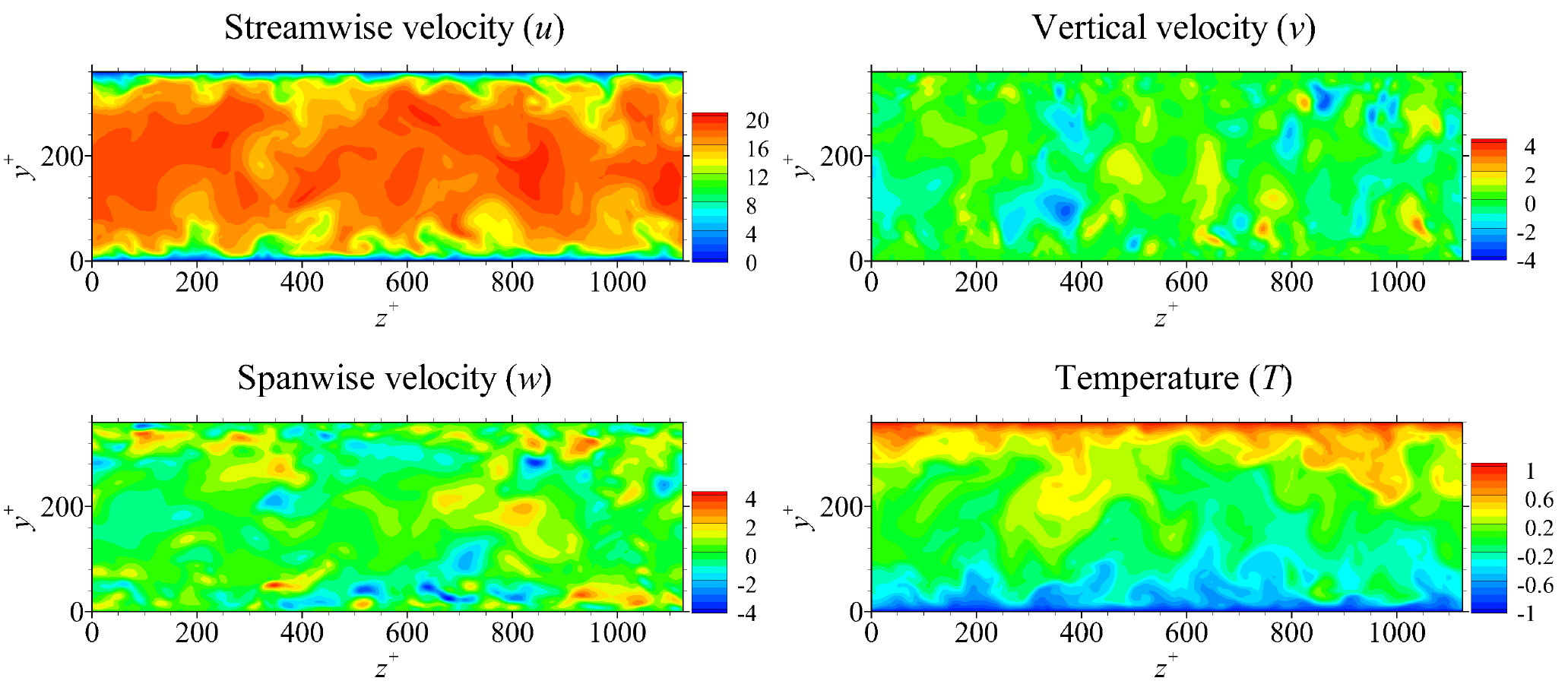}
  }
    \caption{\label{fig:1} The velocity and temperature fields in the y-z plane at $Re_\tau=180$.}
\end{figure}

As shown in Fig. \ref{fig:1}, we collected the data of four variables of flow in the cross-sectional ($y-z$) plane including the streamwise velocity ($u$), vertical velocity ($v$), spanwise velocity ($w$), and temperature ($T$). Through simulations for more than 18 000 wall time units, a total of 20 000 training fields were collected for each Reynolds number. The time interval between temporally successive training fields is 0.9, which corresponds to 10, 20, and 20 time steps of DNSs at $Re_\tau=180, 360,$ and $540$, respectively. The deep learning models performed in this study were trained to generate fields with a time interval of 0.9 ($\Delta t^+_{DL}=0.9$). Because 20 000 collected fields might not be enough for deep learning, a physically reasonable data augmentation was performed. In the channel flow, mirror augmentation, which mirrors the original data symmetrically based on an axis, can be performed for the spanwise and wall normal directions. Using this, we quadrupled the number of data. Also, the spectral augmentation in the spanwise direction, random phase shift, was used. Those data augmentation methods are expected to impose symmetric and homogeneous characteristics of flow on the deep learning network and can be crucial in situations where the amount of collected data is insufficient. As a preprocess, the variables ($u,v,w,T$) of the training data are normalized to have a mean of 0 and standard deviation of 1. The number of $y$-grids in the training data at all Reynolds numbers were matched to 192 whose positions are Chebyshev nodes and are linearly interpolated by the DNS fields. The size of the fields generated by deep learning is fixed to $192\times 192$.

\section{Unsupervised learning for inflow generation}\label{sec:unsup}
\subsection{Generation of Instantaneous flow fields using GAN}{\label{subsec:GAN}}

Before developing a generator of time-varying flow fields, we applied GAN for producing instantaneous flow fields using the collected data. Since \citet{Goodfellow2014} proposed the GAN, several models for generating 2D images based on a convolution operation have been developed rapidly. In general, a GAN consists of two networks, a generator ($G$) and a discriminator ($D$). The generator is a network that receives random noise ($\vec{z}$) in a latent space (low dimension) as input and generates an image through several discrete convolutions and upsamplings. The discriminator is a network that receives the real image or the fake image generated by $G$ as the input and prints out a value per an image through several discrete convolutions and downsamplings. The smaller the output value of $D$, the more likely the generated image is dissimilar to the real one. The training of these two networks can be described by the following min/max problem \cite{Goodfellow2014}:

\begin{equation}
{\min_{\theta}\max_{\psi} \mathbb{E}_{X\sim P_{r}}[\textnormal{log}(D_\psi(X))] + \mathbb{E}_{\vec{z}\sim P_{\vec{z}}}[\textnormal{log}(1-D_\psi(G_\theta(\vec{z})))]}
\end{equation}
where $D$, $G$, $X$, $G_\theta(\vec{z})$, and $\vec{z}$ are the discriminator, generator, real data, generated data, and random noise vector, respectively. The generated data is sometimes denoted as $\tilde{X}$. Here, $\psi$ and $\theta$ are trainable parameters in the discriminator and generator network, respectively. $\psi$ is trained in the direction of decreasing $D_\psi(G_\theta(\vec{z}))$ and increasing $D_\psi(X)$, while $\theta$ is trained in the direction of increasing $D_\psi(G_\theta(\vec{z}))$. That is, $\psi$ is trained in the direction of increase in the difference between the real image and the generated image, and $\theta$ is trained in the direction of catching up the difference. When this adversarial training is performed well, the distribution of data generated by $G$ becomes similar to that of the training data.

The above classical GAN has a non-convergence problem that the learning is oscillatory or a discriminator easily overwhelms a generator so that the learning does not proceed anymore. Also, there is a mode collapse problem that a generator generates only limited varieties of images. To prevent these problems, many attempts have been made to modify the above basic loss function and the architecture. Among them, WGAN-GP loss, which applies a gradient penalty to the discriminator has shown successful performance. It can be described as follows.
\begin{equation}
{L_{D} = \mathbb{E}_{X\sim P_{X}}[-D(X)] + \mathbb{E}_{\tilde{X}\sim P_{\tilde{X}}}[D(G(\vec{z}))] + \gamma \mathbb{E}_{\hat{X}\sim P_{\hat{X}}} [(\| \nabla_{\hat{X}}D(\hat{X}) \|_{2}-1)^2]} \label{WGAN-GP}
\end{equation}
where $\tilde{X}$ and $\hat{X}$ are the generated data ($G(\vec{z})$) and interpolated data, respectively along the straight lines between $X$ and $\tilde{X}$. Here, $\gamma$ is the gradient penalty strength. This loss is defined on the basis of the Wasserstein divergence \cite{Arjovsky2017} instead of the Jensen--Shannon divergence of the original GAN \cite{Goodfellow2014}. The difference between the two models is that the probabilistic divergence of WGAN-GP between the real and generated data distributions becomes continuous with respect to the parameters in the generator by adding the gradient penalty \cite{Gulrajani2017} (last term of Eq. \ref{WGAN-GP}). 

In addition to above approaches, there are several methods of modifying the architecture to improve the generated image quality. For example, \cite{Karras2017} suggested a minibatch standard deviation layer, some normalization techniques, and progressive growing in terms of image resolution. In their study, the use of a minibatch standard deviation layer increased the diversity of the generated image. Also, normalization techniques including equalized learning rate of weights and pixelwise normalization of the feature maps in hidden layers made the training process stable and fast. Progressive growing in terms of image resolution during training can reduce the calculation load and increase the image quality. We tested these methods in our problem. First, we constructed a model that generates instantaneous flow fields without considering the time variation. Unlike the problem of generating a photo image composed of RGB (Red, Green, and Blue) maps, our $G$ was aimed at generating flow fields ($u,v,w,$ and $T$) that have very different shapes on the same spatial location but are correlated with each other. Similarly, $D$ received generated flow fields and real ones. As a result of training, the WGAN-GP loss function produced much better turbulence fields than the classical one. Furthermore, the minibatch standard deviation layer and the normalization techniques improved the training results. However, the use of progressive growing process did not yield noticeable improvement because the resolution of our training data is relatively low, namely $192\times192$. 

For further improvement, we implemented two more approaches, namely using an additional input map on $D$ and applying a statistical constraint on $G$. The first approach is to add the streamwise vorticity component ($\omega_x$) as the additional input of $D$. The idea is related to a key point of synthetic eddy methods \cite{Jarrin2006} that model the eddy structures with a shape function. An eddy is a coherent structure that has spatial and temporal characteristics at the inlet plane, and therefore it needs to be well represented. The $\omega_x$ computed based on $v$ and $w$ is added to $D$ as the input with $u,v,w,$ and $T$. The second approach is to apply a statistical constraint \cite{Wu2019} to the GAN model. Although the discriminator can learn the statistics of the real data and generated data, \citet{Wu2019} reported that adding the difference between those statistics to the original loss function enhances the generator and helps the GANs converge fast. This loss function can be described as follows.

Generator loss:
\begin{equation}
{L_{G} =  \mathbb{E}_{\tilde{X}\sim P_{\tilde{X}}}[-D(\tilde{X})] + \textnormal{MSE}_{G}}
\end{equation}
\begin{equation}
{\textnormal{MSE}_{G} =  \lambda_1 \|S(X)-S(\tilde{X})\|_{2}}
\end{equation}
where $X$ and $\tilde{X}$ are composed of $u,v,w,T,$ and $\omega_x$. $G$ generates only $u,v,w,$ and $T$, while $\omega_x$ is obtained from the generated $u$ and $v$. $\textnormal{MSE}_{G}$ is the statistical constraint, and $\lambda_1$ is the strength. $S(X)$, a statistical quantity, was obtained using the entire training data before learning, while $S(\tilde{X})$ was calculated by the generated data per training iteration. The $\textnormal{MSE}_{G}$ cannot be zero; therefore, a very high strength might make the training results  worse. However, with a little fine-tuning of the strength, the $\textnormal{MSE}_{G}$ showed remarkable improvement in our problem. The effect of $\textnormal{MSE}_{G}$ is given in \ref{app:A}, and the best $\lambda_1$ among the tested cases is $10$. We used the spanwise energy spectrum for $S$ because at high wave numbers, the spectrum shows a noticeable difference between the DNS and the original GAN without the statistical constraint. Here, the energy spectrum $E_{V_iV_i}$ is defined as follows:
\begin{equation}
{E_{V_iV_i}(y,k_z) = {1\over{2\pi}}\int_{-\infty}^{\infty}{e^{-irk_z}R^s_{V_iV_i}(y,r)dr}}
\end{equation}
where 
\begin{equation}
{{R^s_{V_iV_i}(y,r) = {\left<V_i(y,z)V_i(y,z+r) \right>}}}.
\end{equation}
Here, $\left< \right>$ denotes an average operation, and $V_i$ represents the variables including $u,v,w,T,$ and $\omega_x$. Since the spectrum has very different scales depending on the wave number ($k_z$), we used its logarithmic value with a small value ($10^{-10}$) added. This spectrum was calculated by a fast Fourier transform. Furthermore, since the pseudo-spectral method with zero padding was used in our DNS, the energy spectrum of the training data is zero at high wave numbers ($k_z > 64$). To match the statistics of the generated data with the training one, the generated variables ($u,v,w$, and $T$) were zero-padded in high wave numbers. It is a trivial process needed for the case using the pseudo-spectral method, and other simulation data might not need this process. The calculation time per training iteration of the GAN with $\textnormal{MSE}_{G}$ is almost the same as that of the original GAN without it, and the implementation is not difficult. Therefore, we recommend to use this statistical constraint \cite{Wu2019} in the problem that applies the GAN to the turbulence research. Especially, when particular statistics are important in some problems, those can be prioritized to fit well to those of the real data.

\begin{figure}
  \centerline{
  \includegraphics[width=0.9\columnwidth]{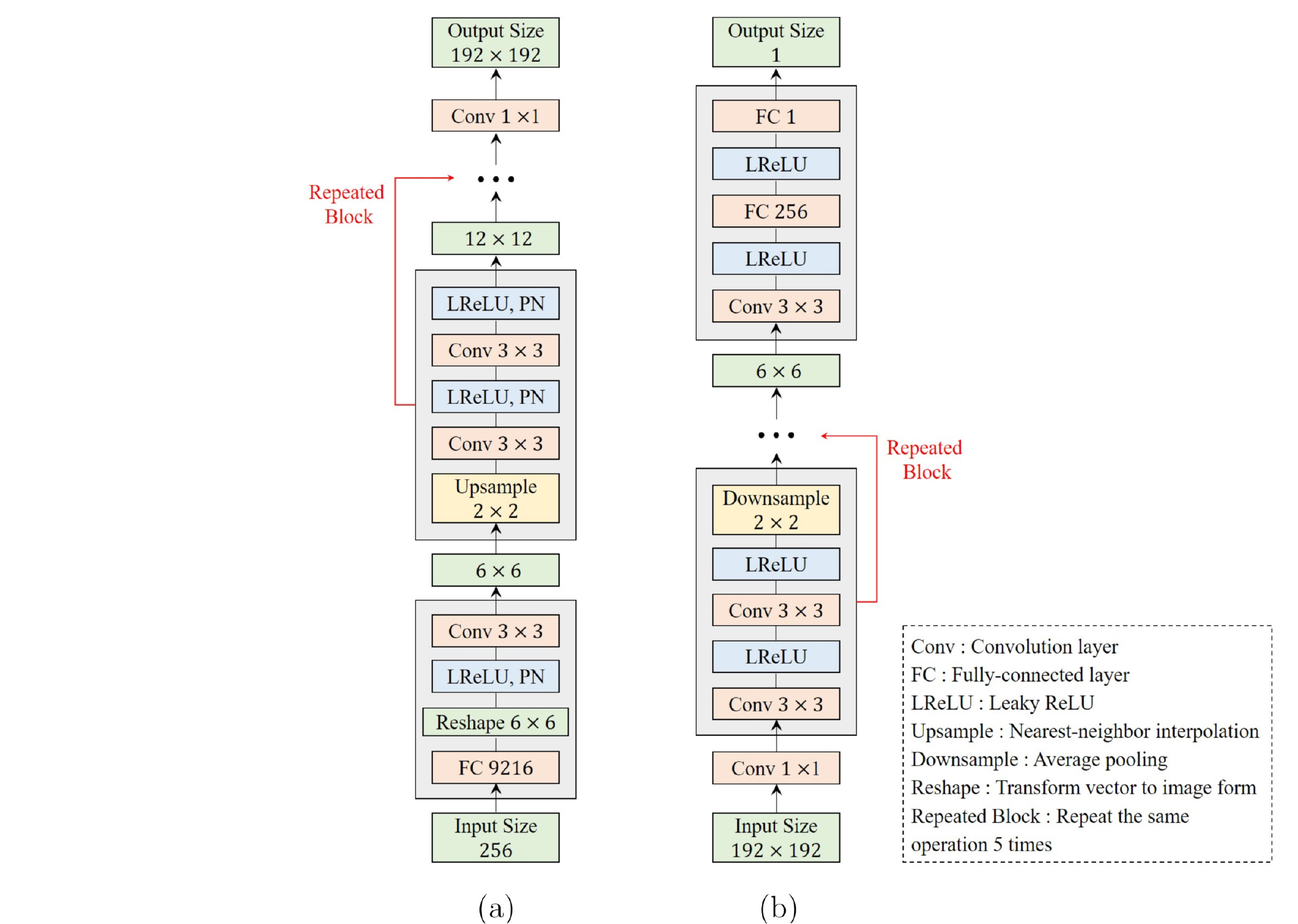}
  }
  \caption{\label{fig:2}{Architecture of (a) generator and (b) discriminator networks.}}
\end{figure}

We trained the WGAN-GP using the instantaneous flow fields in the $y-z$ plane of the turbulent channel flow.  Fig. \ref{fig:2} shows the detailed network architecture consisting of the generator and discriminator, which is similar to that of \citet{Karras2017}. They are constructed by blocks composed of two convolution layers and upsampling (or downsampling). Also, $3\times 3$ kernel is used for most convolution layers. Periodic padding is applied to the spanwise direction of the convolution operation in the generator and discriminator networks so that the periodic boundary condition is naturally satisfied. In addition, $2\times2$ average pooling and nearest-neighbor interpolation is used for downsampling and upsampling, respectively. As a nonlinear function, the leaky ReLU function most used in GAN is applied. The number of input nodes in the latent space is 256, and each node is sampled from a normal distribution. For the training, 20 000 fields at $Re_\tau=180$ were used. Mirror-augmentation and spectral augmentation were used to increase the number of data. These augmentation methods made the training stable and the statistics converged better; their deails are presented in \ref{app:B}. Adaptive moment estimation (ADAM) \cite{Kingma2014}, one of the gradient descent methods, was used for the training with a fixed learning rate of 0.001, and the training was carried out for a total of 300 000 iterations with a batch size of 8 per iteration. More training did not yield a significant quality-improvement in the generated fields. This method was implemented using the open source library TensorFlow \cite{Abadi2016}.

\begin{figure}
  \centerline{
  \includegraphics[width=1.0\columnwidth]{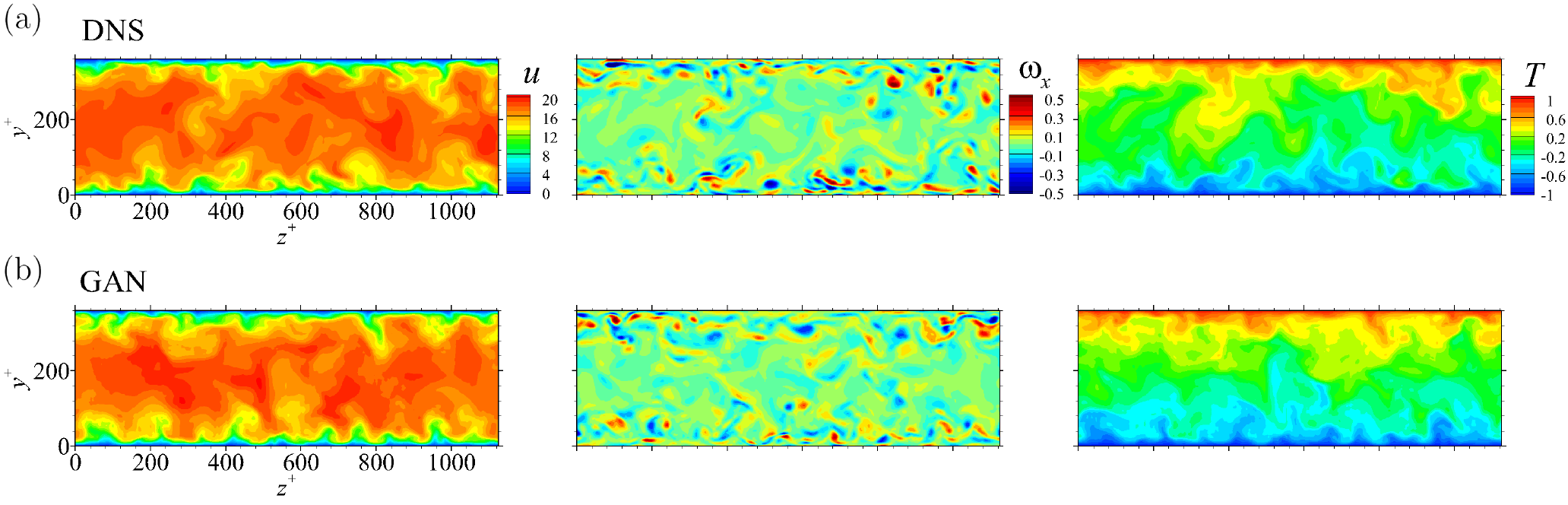}
  }
  \caption{\label{fig:3}{Generated fields at $Re_\tau=180$. The streamwise vorticity is normalized by $Re_\tau$.}}
\end{figure}

\begin{figure}
  \centerline{
  \includegraphics[width=0.9\columnwidth]{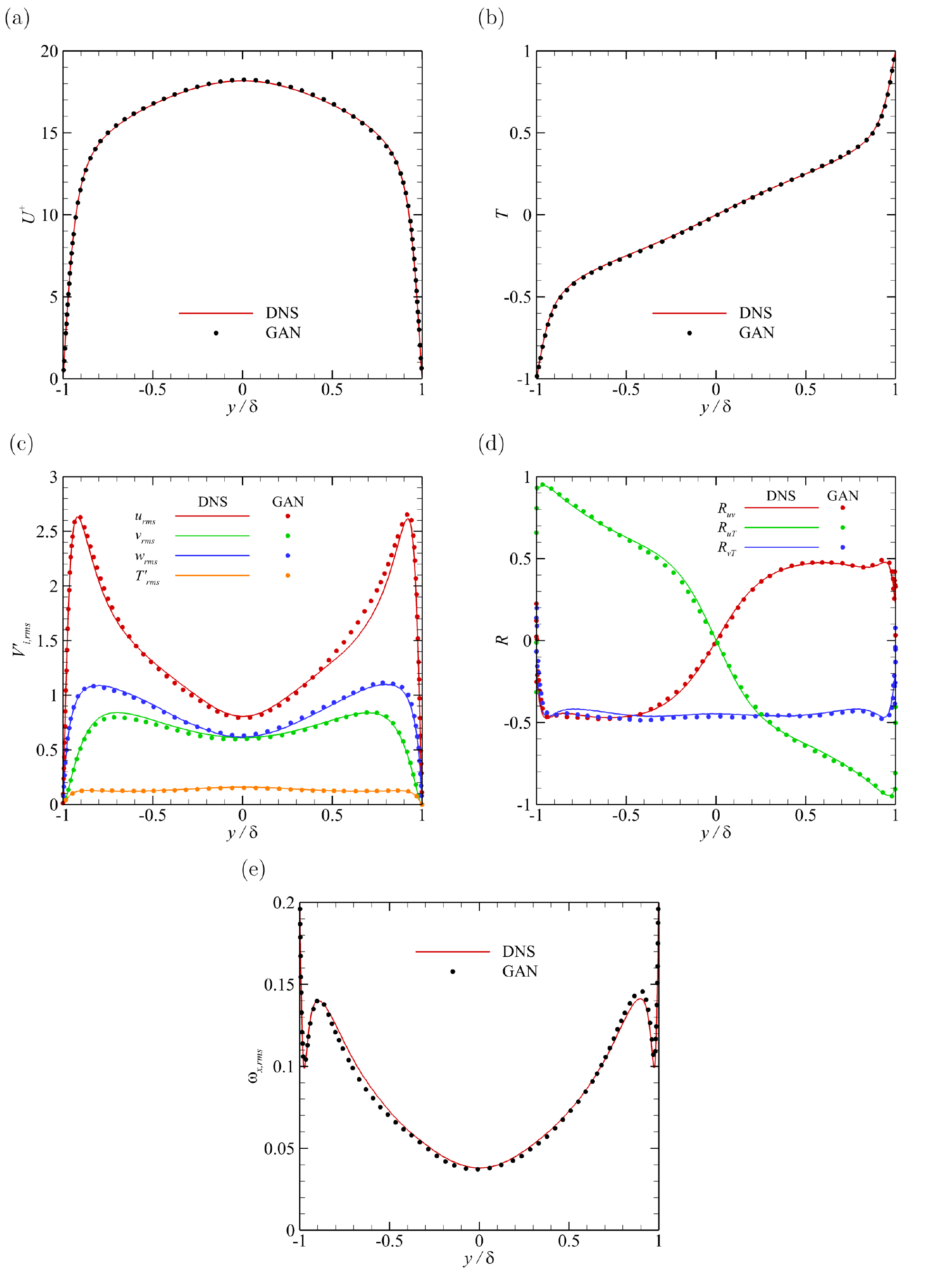}
  }
  \caption{\label{fig:4}{Statistics of fields generated by GAN at $Re_\tau=180$. (a) and (b) show the streamwise mean velocity and mean temperature profiles, respectively. (c) RMS profiles of the trained variables. (d)  Pointwise correlations. (e) RMS profiles of the streamwise vorticity, which is obtained by the vertical and spanwise velocities.}}
\end{figure}

After training, an instantaneous flow field can be generated using only $G$, 
\begin{equation}
{\tilde{X} = G(\vec{z})}
\end{equation}
where $\vec{z}$ is a random noise with a normal distribution, and $\tilde{X}$ includes $u,v,w,$ and $T$. A turbulence field generated by the trained generator at $Re_\tau=180$ is shown along with a DNS field in Fig. \ref{fig:3}. The present GAN model expresses the small-scale structures very well. It is not easy to distinguish between the fields generated by the GAN and the DNS. In addition, the basic statistics of the generated fields, including the mean, root-mean-square (rms) of fluctuations, and pointwise correlation, are compared with those of the DNS in Fig. \ref{fig:4}. Here, the pointwise correlation is defined as follows:
\begin{equation}
{R_{V_iV_j}(y) = {cov(V_i(y),V_j(y)) \over \sigma_{V_i(y)}\sigma_{V_j(y)}}}
\end{equation}
where $cov$, $\sigma$, and $V_j$ are the covariance, standard deviation, and variables ($u,v,w,$ and $T$), respectively. The statistics of the GAN were ensemble-averaged using 10 000 independently generated fields. Those of the DNS were calculated using the training data. The streamwise mean velocity and temperature profiles of the GAN are almost the same as those of DNS. The rms profiles of the GAN are in very good agreement with those of the DNS. Detailed distributions of the correlations, $R_{uv}$, $R_{uT}$, and $R_{vT}$, of the DNS data are very well captured by the fields generated by GAN. It indicates that in the training of GAN, the correlations between different variables have been well considered. By symmetry, the correlations such as $R_{uw}$, $R_{vw}$, and $R_{wT}$ of the generated fields are zero (not shown here). The rms of the streamwise voriticty normalized by $Re_\tau$ is accurately simulated by GAN. This successful performance of GAN clearly indicates that unsupervised learning of turbulence is indeed possible. 

Next, we extended this model to a more useful one that could generate instantaneous flow fields at various Reynolds numbers. In order to implement the effect of the Reynolds number in the generation network, we can take advantage of the latent space of GAN. Typical GAN has a noticeable characteristic that a linear variation of $\vec{z}$, which is the input of the generator, can result in a meaningful change in the generated image \cite{Radford2015,Berthelot2017}. For example, when $\vec{z}_1$ and $\vec{z}_2$ generate face images looking at the left and the right, respectively, $\vec{z}_1+a(\vec{z}_2-\vec{z}_1)$ with $a$ in the range $0<a<1$ could be the vector that produces the face image looking at a specific direction between left and right. 
Using this function, we investigated whether a linear variation of a component in latent space ($\vec{z}$) can be linked to a change in the Reynolds number in a flow field produced by the generator. Specifically, we inserted a Reynolds number instead of random noise in a component ($z_1$) where $\vec{z}=(z_1,z_2,..)$, and then we expected that the generated image ($G(\vec{z})$) becomes a flow field at this Reynolds number. For this, we need a guide so that $z_1$ indicates the Reynolds number of the generated field. Unlike the previous discriminator that receives only flow fields ($u,v,w,T,$ and $\omega_x$) as input, we added another channel filled with the pixel value of the Reynolds number ($z_1$ value) to the input of the discriminator. We call this channel $Re$ map. Namely, the discriminator receives six channels including $u,v,w,T,\omega_x,$ and $Re$. More specifically, to extract the universal characteristics of the near-wall turbulence, we filled the $Re$ map with the wall coordinates, which contains Reynolds number information.

\begin{figure}
	\centerline{
		\includegraphics[width=1.0\columnwidth]{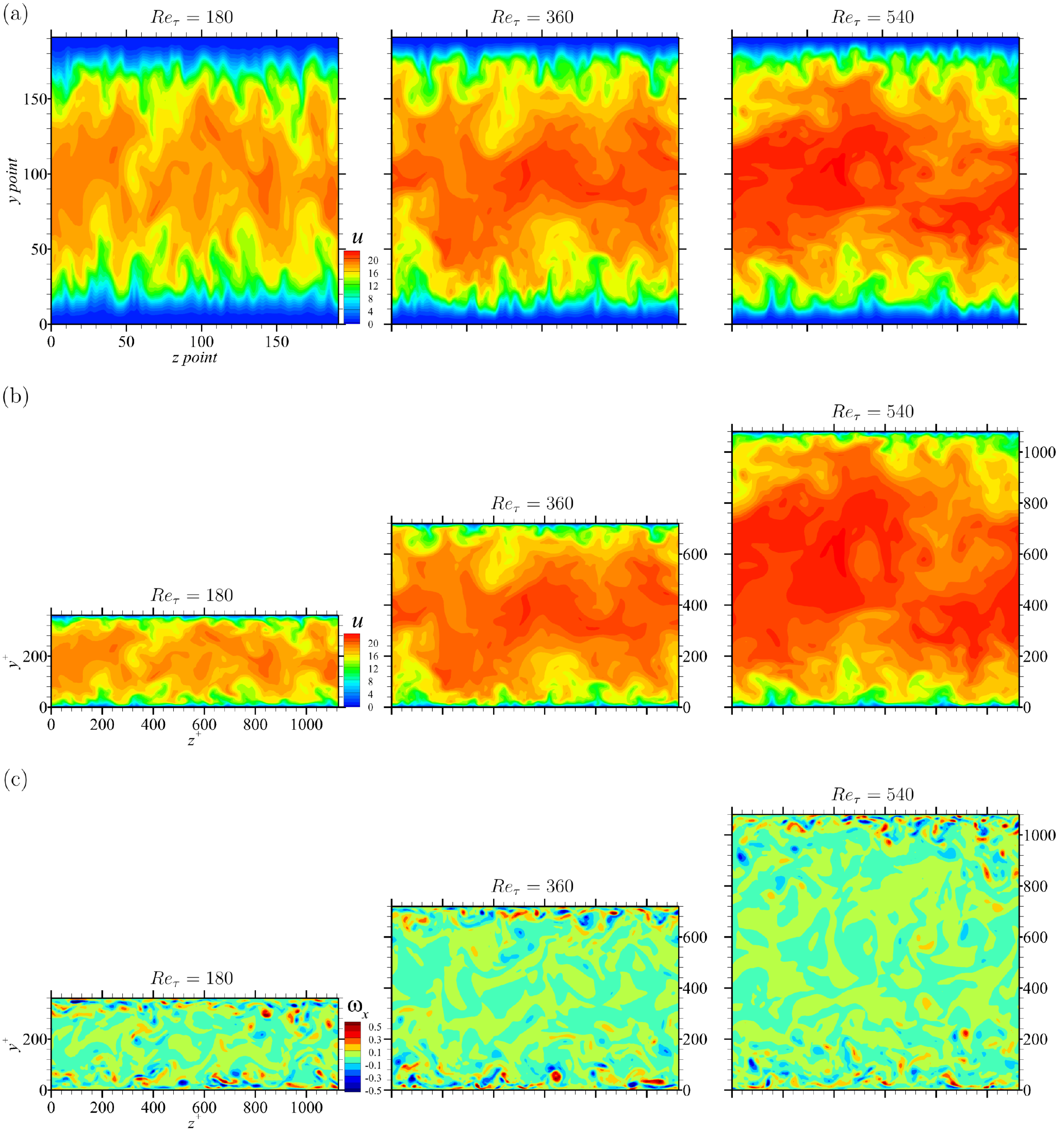}
	}
	\caption{\label{fig:5}{Collected data at three Reynolds numbers. (a) Streamwise velocity fields with grid index coordinates. (b) Streamwise velocity fields with physical wall coordinates. (c) Streamwise vorticity fields with physical wall coordinates. The vorticity is normalized by $Re_\tau$.}}
\end{figure}

We trained a GAN using flow fields at three Reynolds numbers ($Re_\tau=180, 360, 540$). The size of the training data was fixed at $192\times 192$ for all three Reynolds numbers. Samples of the collected data at three Reynolds number are shown in Fig. \ref{fig:5}(a), in which the fields are illustrated in terms of the grid indices, not the physical coordinates. Our GAN learns these kinds of images in which the Reynolds number effect is represented in the grid index spaces. Figs. \ref{fig:5}(b) and (c) have the same fields in the physical wall coordinates. Near-wall turbulent structures, such as low- and high- speed streaks and vortical motions, show similar characteristics regardless of the Reynolds number. Of course, the turbulent intensity in the wall unit slightly increases with the Reynolds number. Therefore, we tested whether our GAN can capture the Reynolds number effect observed in Figs. \ref{fig:5}(b) and (c), after learning the data given in the form shown in Fig. \ref{fig:5}(a).

\begin{figure}
	\centerline{
		\includegraphics[width=1.0\columnwidth]{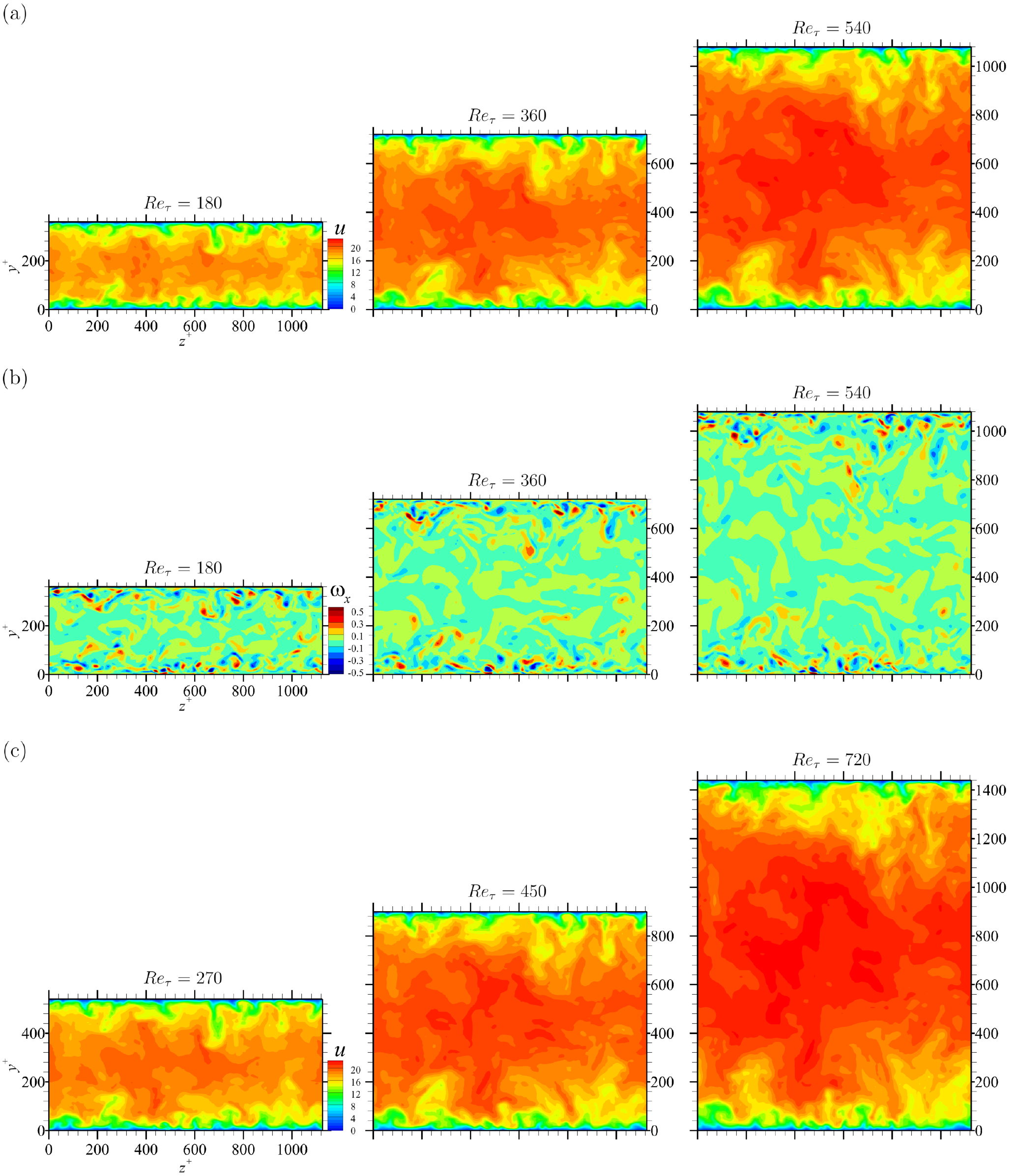}
	}
	\caption{\label{fig:6}{The generated fields by GAN at several Reynolds numbers. (a) and (b) are the streamwise velocity and vorticity fields normalized by $Re_\tau$ at the trained Reynolds numbers ($Re_\tau = 180, 360, 540$), respectively. (c) Streamwise velocity fields at the new Reynolds numbers ($Re_\tau =  270, 450, 720$), which are not trained.}}
\end{figure}

\begin{figure}
	\centerline{
		\includegraphics[width=1.0\columnwidth]{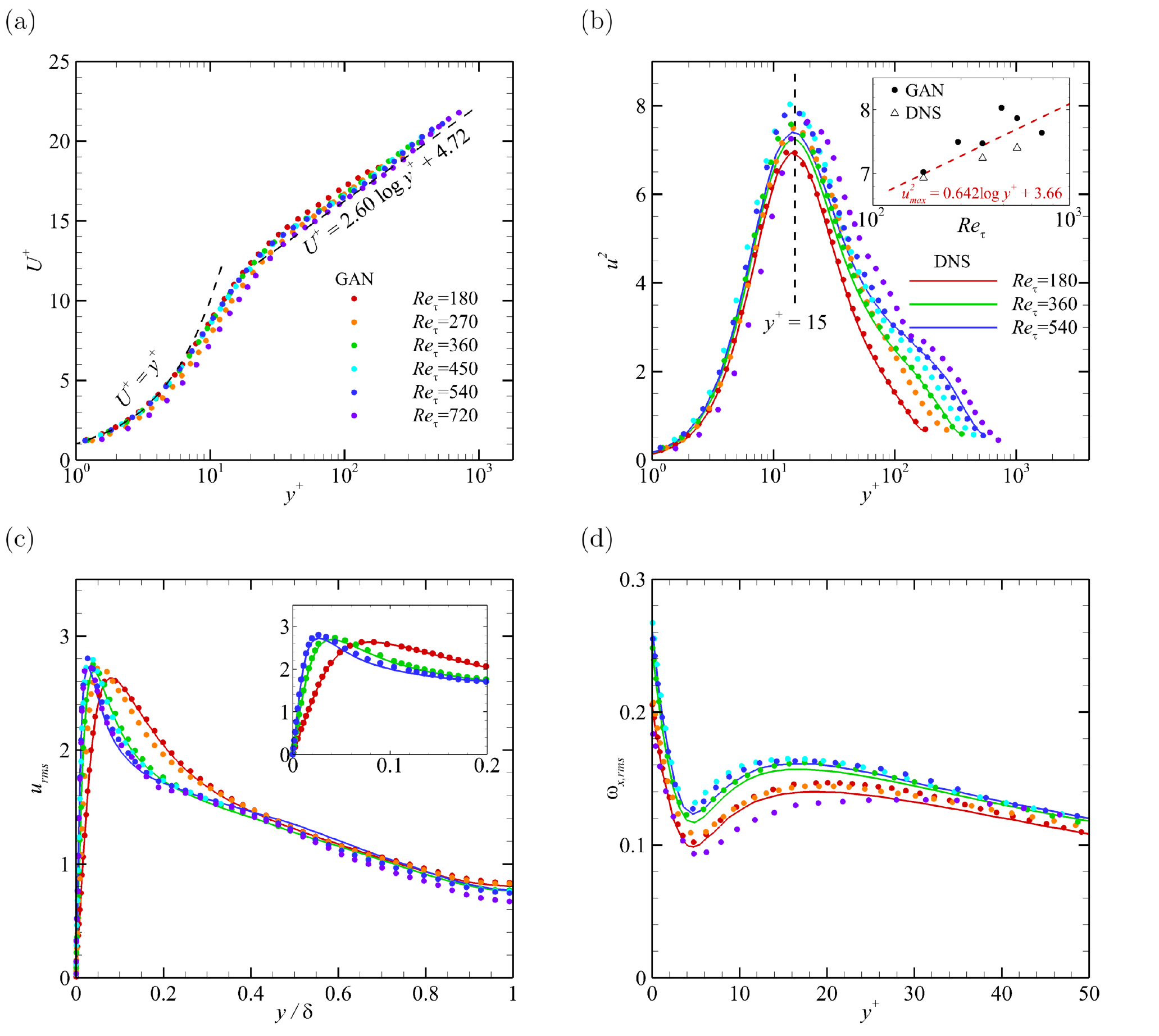}
	}
	\caption{\label{fig:7}{Statistics of the fields generated by GAN at several Reynolds numbers. (a) Streamwise mean velocity with wall coordinates. (b) Mean square profiles of streamwise velocity fluctuations with wall coordinates, and the maximum values with respect to $Re_\tau$ obtained with the equation of \citet{Lee2015} as shown in the inset figure. (c) RMS profiles of streamwise velocity fluctuations with global coordinates; the inset figure is a magnified view near the wall. (d) RMS profiles of streamwise vorticity normalized by $Re_\tau$. Data at $Re_\tau = 180, 360, 540$ is trained; however, data at $Re_\tau = 270, 450, 720$ is not trained.}}
\end{figure}

Training was carried out with the same hyperparameters as those of the GAN learning with only single Reynolds number data except for the batch size. The batch size was set to 6 with 2 for each Reynolds number. After training, the flow fields were generated at several Reynolds numbers using the trained generator for the same $\vec{z}$ except the first component, and they are shown in Fig. \ref{fig:6}. At the trained Reynolds numbers ($Re_\tau=180, 360, 540$), the generated fields (Fig. \ref{fig:6}(a)) are very similar to the DNS fields (Fig. \ref{fig:5}(b,c)) in that the turbulent structures near the wall, such as vortical motions, are similar in size and strength. Even at new Reynolds numbers ($Re_\tau=270,450,720$, see Fig. \ref{fig:6}(c)), the generated fields  show similar low-speed structures near the wall.  An interesting observation from Fig. \ref{fig:6} is that the GAN generates almost the same turbulent structures in the wall coordinates in spite of the changes in the Reynolds number component in $\vec{z}$. It is inferred that the GAN is trained to the direction of minimizing the change in the near-wall structures according to the change in the Reynolds number, in the same spanwise domain size in wall units as that of the trained Reynolds numbers. From this, we can conclude that a successful unsupervised learning of turbulence results from the existence of the near-wall turbulent structures, especially vortical structures, which have similar shape in wall units relatively insensitive to the Reynolds number.  

The detailed statistics of the generated fields at various Reynolds numbers are also compared against the DNS data. The statistics were acquired by taking ensemble average over 10 000 generated fields at each Reynolds number. As shown in Fig. \ref{fig:7}(a), the mean velocity profiles of the generated fields within the trained range ($180\leq Re_\tau \leq540$) follow very well the linear profile in the viscous sublayer and the log profile ($U^+=2.60~ \textnormal{log}(Re_\tau)+4.72$ \cite{Lee2015}) in the logarithmic region, while a slight deviation is observed for $Re_\tau =720$. In Fig. \ref{fig:7}(b), the mean-square velocity profiles of the generated fields also show a pick value at $y^+\approx15$ regardless of the Reynolds number. The maximum value, however, is slightly scattered around the log function with respect to the frictional Reynolds number, $u^2_{max}=0.642\textnormal{log}(Re_\tau)+3.66$ \cite{Lee2015}. It might be related to the very small difference in the maximum value at the trained Reynolds numbers. When the Reynolds number exceeds the trained one ($Re_\tau>540$), both the mean and the mean-square velocity profiles tend to be inaccurate. The root-mean-squared profiles of the streamwise velocity in the outer variable are shown in Fig. \ref{fig:7}(c). Far from the wall, $y > 0.4\delta$, the profiles including the DNS ones almost collapse within the trained Reynolds number range, as reported by \cite{Moser1999}. Likewise, at $Re_\tau>540$, the profile slightly deviates from the collapsed profile. Near the wall, the rms profiles of the generated fields are well matched with those of the DNS at the trained Reynolds numbers. Because the GAN learned the data in the grid index coordinates (Fig. \ref{fig:5}(a)), the characteristic that has a different pick position in the outer variable at each Reynolds number is not easy to learn. Nevertheless, the GAN shows remarkable learning ability and the potential to be applied to turbulence simulations. In Fig. \ref{fig:7}(d), the rms profiles of vorticity, normalized by $Re_\tau$, are in fairly good agreement with those of DNS and show an increasing tendency with the Reynolds number except the case beyond the trained range.

Our tests indicate that the GAN could learn the universal nature of the Reynolds number effect by capturing the similarity hidden in the chaotic turbulence data through the latent space, $\vec{z}$, suggesting the possibility of reflecting other effects or parameters through the latent space. So far, we have only dealt with the problem of generating the instantaneous flow fields, naturally leading to discussion of the problem of generating time-varying flow fields.

\subsection{RNN-GAN for time-varying flow generation}\label{subsec:RNN-GAN}

In this section, we construct a GAN to represent the time-variation so that the trained GAN can generate time-varying turbulence fields. The basic idea is similar to that of reflecting the Reynolds number effect on the GAN. It is assumed that a flow change in time could be performed by changing the $\vec{z}$ in the latent space, which is the input of the trained generator. We aim to obtain $\vec{z}_{t+1}$, which can generate the flow information at the next time step, by advancing $\vec{z}_{t}$ in a low dimension. Here, IndyLSTM \cite{Li2018,Gonnet2019}, a kind of recurrent neural networks (RNNs), is used to update $\vec{z}_{t}$ to $\vec{z}_{t+1}$. A similar approach that applies RNNs to the time advancement in a low dimension was used for the prediction of unsteady flows \cite{Wang2017,Mohan2018,Mohan2019}. Because only a small amount of compressed information is used, there is an advantage that the time update part can be performed with only a small amount of memory and operation.

The RNN is a model that efficiently handles sequential data and can predict the next step information based on the previous step information and input information. Using the notation of \citet{Li2018}, the RNN can be expressed as follows:
\begin{equation}
{h_{t} = \sigma (Wx_{t}+Uh_{t-1}+b)}
\end{equation}
where $x_{t}$ and $h_{t}$ are the input and output at a time step $t$, respectively. Also, $h_{t}$ could be the recurrent input at a time step $t+1$. $W, U$, and $b$ are the weights for the current input, the weights for the recurrent input, and the bias, respectively, and $\sigma$ is an elementwise activation function. The RNN has a characteristic that the time length of input and output information is variable. In other words, even if the training time-length is 500 steps, information beyond 500 steps might be predicted after training. Of course, the accuracy of prediction beyond the trained range is not guaranteed as much as that of the trained range. The inflow generator that we are trying to build should guarantee this accuracy for a very long time, and the method to achieve this is discussed later. IndyLSTM, which is a model developed from the basic RNN, is used in our problem and can be described as follows:

\begin{equation}
{f_{t} = \sigma_g (W_{f}x_{t}+U_{f}\odot h_{t-1}+b_f)}
\end{equation}
\begin{equation}
{i_{t} = \sigma_g (W_{i}x_{t}+U_{i}\odot h_{t-1}+b_i)}
\end{equation}
\begin{equation}
{o_{t} = \sigma_g (W_{o}x_{t}+U_{o}\odot h_{t-1}+b_o)}
\end{equation}
\begin{equation}
{c_{t} = f_t \odot  c_{t-1} + i_t \odot  \sigma_c (W_{c}x_{t}+U_{c}\odot h_{t-1}+b_c)}
\end{equation}
\begin{equation}
{h_{t} = o_t \odot  \sigma_h (c_{t})}
\end{equation}
where $x_{t}$ is the current input, and $c_{t}$ and $h_{t}$ are respectively the hidden state and cell state, which store the information of the previous step. $f_ {t}$, $i_{t}$, and $o_{t}$ are the forget gate, input gate, and output gate, which determine how much to forget about the previous state, how much input is received, and how much to output, respectively. These are determined by the trainable parameters $W$, $U$, and $b$, respectively. The sigmoid function is mostly used for $\sigma_g$, and the tanh function is mostly used for $\sigma_c$ and $\sigma_h$. The difference between the classical LSTM and IndyLSTM is that IndyLSTM changes the fully-connected nature of $h_{t}$ to elementwise product (Hadamard product, $\odot$) for the purpose of preventing gradient vanishing and exploding. Although the classical LSTM also guarantees the gradient to a certain extent through an addition operation between the previous step and the next step, IndyLSTM showed better performance than the traditional LSTM in our test.

\begin{figure}
  \centerline{
  \includegraphics[width=1.0\columnwidth]{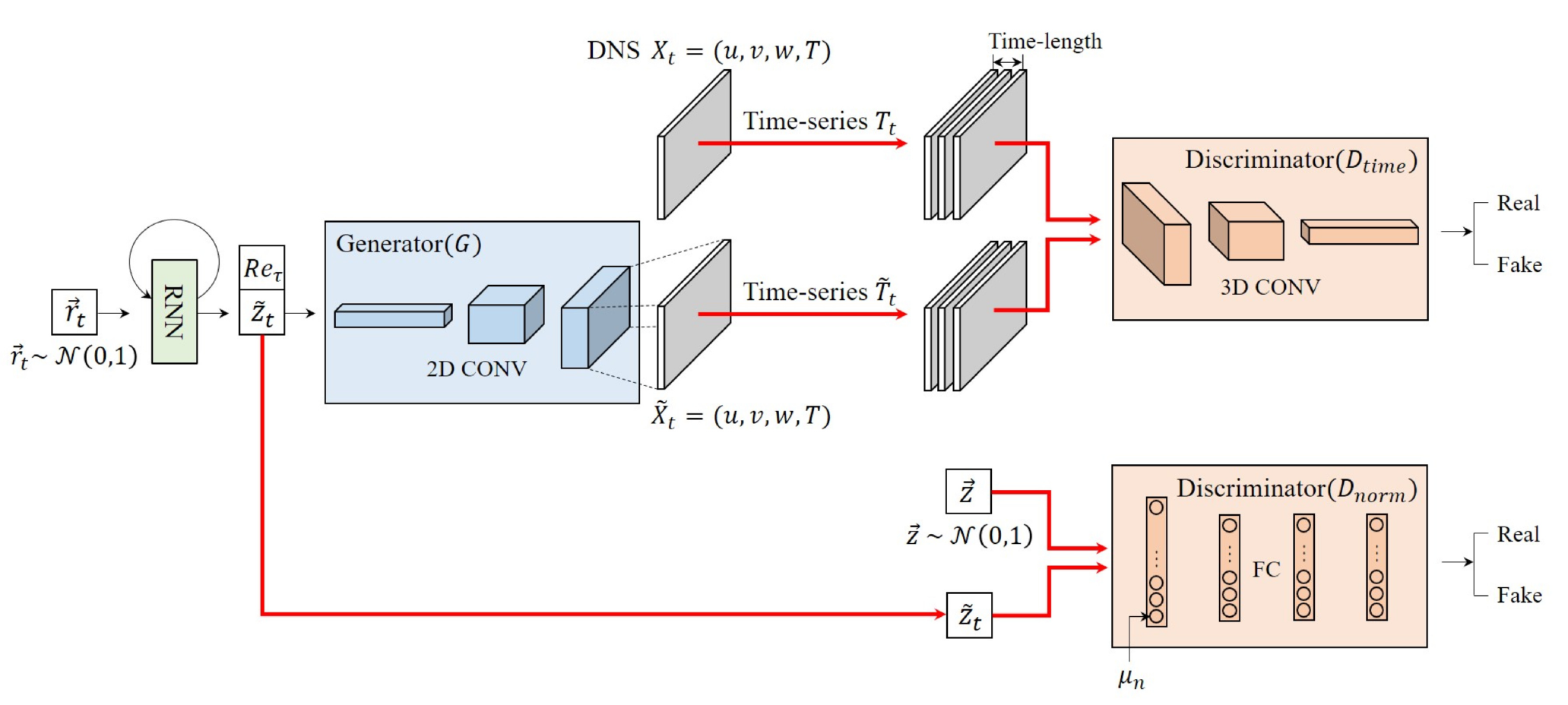}
  }
  \caption{\label{fig:8}{Schematic diagram of RNN-GAN.}}
\end{figure}

 In the inflow generation problem, the stochastic time-variation of flow is required. Unlike the supervised learning method, our unsupervised learning method can reflect this stochastically varying condition at the inlet boundary by inserting random noise ($\vec{r}_{t}$) as the input ($x_{t}$) of the RNN. The $\vec{r}_{t}$ is set to have a normal distribution. As shown in Fig. \ref{fig:8}, we expected to obtain $\tilde{z}_{t}$ that can produce a statistically valid flow field through the pretrained generator network ($G$), putting the random noise sequentially ($\vec{r}_{1},...,\vec{r}_{t}$). Also, we expected that the time-sequential flow fields ($\tilde{X}_{t},\tilde{X}_{t+1},\tilde{X}_{t+2},...$) are produced from the sequential outputs ($\tilde{z}_{t},\tilde{z}_{t+1},\tilde{z}_{t+2},...$) of the RNN. In order to achieve this goal, a new discriminator $D_{time}$ (Fig. \ref{fig:8}) was constructed for training the RNN, which receives the time-sequential flow fields as input. If we take the three sequential flow fields over time as the input of the $D_{time}$, the input size becomes $3\times 192\times 192\times 6$ (t-length $\times$ y-grids $\times$ z-grids $\times$ channels). In the $D_{time}$, 3D convolution layers considering the time-directional convolution operation were used for efficiently extracting the time-variation of flow.

Before training the RNN-GAN, which is expected to time-dependently produce a fully developed flow, there is something to be concerned about RNN, as mentioned above. Because the generated flow should be statistically valid for a very long period of time, the output of the RNN must be statistically stationary. If training is performed only for short steps, say from $0$ to $10$, there is no guarantee that the trained RNN generates statistically stationary output after 10 steps. In particular, it is known that the data generated by the RNN at early steps (for example, $0\sim10$ steps) is statistically transient \cite{Cooijmans2016}. First, to generate a statistically stationary output, we set the length ($t_{max}$) of the RNN to be as long as 500 steps during training. Second, we tried to make the output ($\tilde{z}_{t}$) of RNN be a normal distribution, because the pretrained $G$ can generate a statistically accurate field from the vector with a normal distribution. To implement this, we constructed one more discriminator $D_{norm}$ that distinguishes whether $\tilde{z}_{t}$ at a randomly selected step is the normal distribution (Fig. \ref{fig:8}). Not only $\tilde{z}_{t}$ but also the n-th order moments ($\mu_n$) of $\tilde{z}_{t}$, including the mean, rms, skewness, and flatness, were used as the input of the $D_{norm}$ to distinguish the distribution more efficiently.

When training only the parameters in the RNN, a typical problem is that the temporal correlations of all variables ($u,v,w,T$) decrease over time at a similar rate. The reason is that the diversity of instantaneous flow fields that the pretrained generator ($G$) can produce is not strong enough to perfectly generate a time-varying flow. Therefore, we also trained the parameters in the $G$ in addition to those in the RNN. In summary, the RNN-GAN model, which could generate a statistically stationary turbulent flow for a long time, consists of $G$, $RNN$, $D$, $D_{time}$, and $D_{norm}$. Using two or more discriminators is not a new idea. \citet{Xie2018} also used spatial and temporal discriminators simultaneously for a supervised learning problem. Also, the idea of separating the temporal and spatial representations in the generator was used by \citet{Saito2017}. The major difference is that we use the RNN to generate data for a very long period and do not fix a generated time-length. Our loss functions for training the model can be described as follows.

Discriminator loss :
\begin{equation}
{L_{D} =\mathbb{E}_{X\sim P_{X}}[-D(X)] + \mathbb{E}_{\tilde{X}\sim P_{\tilde{X}}}[D(\tilde{X})] + \gamma_{1}  \mathbb{E}_{\hat{X}\sim P_{\hat{X}}} [(\| \nabla_{\hat{X}}D(\hat{X}) \|_{2}-1)^2]}
\end{equation}
\begin{equation}
{L_{D_{time}} = \mathbb{E}_{T_{t}\sim P_{T_{t}}}[-D_{time}(T_{t})] + \mathbb{E}_{\tilde{T}_{t}\sim P_{\tilde{T}_{t}}}[D_{time}(\tilde{T}_{t})] + \gamma_{2}  \mathbb{E}_{\hat{T}_{t}\sim P_{\hat{T}_{t}}} [(\| \nabla_{\hat{T}_{t}}D_{time}(\hat{T}_{t}) \|_{2}-1)^2]}
\end{equation}
\begin{equation}
{L_{D_{norm}} = \mathbb{E}_{\vec{z}\sim P_{\vec{z}}}[-D_{norm}(\vec{z})] + \mathbb{E}_{\tilde{z}_t\sim P_{\tilde{z}_t}}[D_{norm}(\tilde{z}_t)]} 
\end{equation}

Generator loss :
\begin{equation}
{L_{G} = \mathbb{E}_{\tilde{X}\sim P_{\tilde{X}}}[-D(\tilde{X})] + \mathbb{E}_{T_{t}\sim P_{T_{t}}}[-D_{time}(T_{t})] + {\textnormal{MSE}}_{G} + {\textnormal{MSE}}_{RNN\textnormal{-}G}}
\end{equation}
\begin{equation}
{L_{RNN} = \mathbb{E}_{\tilde{T}_{t}\sim P_{\tilde{T}_{t}}}[-D_{time}(\tilde{T}_{t})] + \mathbb{E}_{\tilde{z}_t\sim P_{\tilde{z}_t}}[-D_{norm}(\tilde{z}_t)] + {\textnormal{MSE}}_{RNN\textnormal{-}G} + {\textnormal{MSE}}_{RNN}}
\end{equation}
where
\begin{equation}
{\textnormal{MSE}_{G} = \lambda_{1} \|S_1(X)-S_1(\tilde{X})\|_{2}}
\end{equation}
\begin{equation}
{\textnormal{MSE}_{RNN\textnormal{-}G} = \lambda_{1} \|S_1(T_t)-S_1(\tilde{T}_t)\|_{2} + \lambda_{2} \|S_2(T_t)-S_2(\tilde{T}_t)\|_{2}}
\end{equation}
\begin{equation}
{\textnormal{MSE}_{RNN} = \lambda_{3} \|S_3(\vec{z})-S_3(\tilde{z}_t)\|_{2} + \lambda_{4} \sum_k k \| R_{\tilde{z}_t\tilde{z}_{t+k}} \|_{2}}
\end{equation}
where $L_D$, $L_{D_{time}}$, $L_{D_{norm}}$, $L_{G}$, and $L_{RNN}$ are the loss functions for training $D$, $D_{time}$, $D_{norm}$, $G$, and $RNN$, respectively. Here, $D$ and $G$ denotes the discriminator and generator used in the instantaneous field generation, respectively. Also, $X$, $\tilde{X}$, $T_t$, $\tilde{T}_{t}$, $\vec{z}$, and $\tilde{z}$ denote the DNS field, the generated field, the time series of DNS fields, the time series of generated fields, random noise vector, and the randomly sampled vector among the outputs of RNN. The time-series fields are $T_{t} = (X_{t},X_{t+1},...)$ and $\tilde{T}_{t} = (\tilde{X}_{t},\tilde{X}_{t+1},...)$. $\hat{T}_{t}$ is the interpolated data between $T_t$ and $\tilde{T}_{t}$. 

The loss with respect to $D$ was used to make the pretrained generator ($G$) produce the instantaneous flow fields with accurate spatial correlations. The loss with regard to $D_{time}$ was used to make $RNN$-$G$ be able to implement the time variation of flow. The loss with respect to $D_{norm}$ was used to maintain $\tilde{z}$, the normal distribution regardless of time. Also, there are $\textnormal{MSEs}$, statistical constraints, to enhance the quality of the generated fields and to increase the convergence speed of training. First, $\textnormal{MSE}_{G}$ was used to represent the spatial characteristics of turbulence well. Here, $S_1$ is the energy spectrum of velocities, temperature, and vorticity. Second, $\textnormal{MSE}_{RNN\textnormal{-}G}$ was adopted to express the time-variation of flow well. $S_2$ represents the temporal correlations of variables ($V_i$) including velocities, temperature, and vorticity. They are defined as follows:
\begin{equation}
{R^t_{V_iV_i}(s) = {cov(V_i(t),V_i(t+s)) \over \sigma_{V_i(t)}^2}.}
\end{equation} Lastly, $\textnormal{MSE}_{RNN}$ was used to ensure that $\tilde{z}_t$ is normally distributed. $S_3$ is the n-th order moment whose expectations are the given constants for the normal distribution. The second term of $\textnormal{MSE}_{RNN}$ is a guide to make the generated flow decorrelated in time. The correlation between the two output vectors of RNN, which are separated by $k$ steps from each other, might  decrease with an increase in the distance. Therefore, we weighted the correlation loss according to the distance, and there is room for improvement through some modifications. Before training, all the statistics of real data is precalculated using the entire training data instead of using batch datasets, while those of the generated data were calculated during training.

As mentioned above, we performed transfer learning that uses the parameters of a pretrained network as the initial parameters. The transfer learning of the pretrained $G$ should be carried out carefully. Otherwise, the sudden deterioration of the pretrained $G$ occurs. For example, the network generates a blurred flow field or highly correlated flow fields from the decorrelated $z$ vectors. The reason is that $D_{time}$ is not yet trained enough in the early training period. These problems could be resolved by the presence of $D$ and the statistical constraints. They maintain the spatial correlation of the generated flow fields by suppressing the sudden deterioration well. Here, we set $\lambda_{1}$, $\lambda_{2}$, $\lambda_{3}$, and $\lambda_{4}$ to 10, 1000, 100, and 1, respectively, considering the magnitude of each loss. The sophisticated tuning of those values might improve training. 

In addition to $\lambda_{i}$, the hyperparameters used in the training of RNN-GAN are as follows. The time interval between $X_{t}^{DNS}$ and $X_{t+1}^{DNS}$ is $0.9$ in wall time units, and a total of 20 000 training fields at each Reynolds number were used for training. We set the 8 step information of $T_{t}=({X}_{t},{X}_{t+1},{X}_{t+3},{X}_{t+7},{X}_{t+15},{X}_{t+31},{X}_{t+63},{X}_{t+63+k})$ as the input of $D_{time}$ to ensure that both short and long time correlations were well represented. At the same time, for the purpose of preventing the network from learning spurious periodicity of DNS data, ${X}_{t+63+k}$ was added. ${X}_{t+63+k}$ is a randomly selected real data that is decorrelated with $X_{t}$, whereas $\tilde{X}_{t+63+k}$ is the generated data at a time step $t+63+k$, where $k$ is a random number larger than $100$ ($\Delta t^+=90$). The number of layers and nodes in the RNN are 3 and 255, respectively. Finally, a linear fully-connected layer is attached to the output of the RNN. Although an increase in the number of layers might improve the training results, we did not investigate it further. The length ($t_{max}$) of the RNN was fixed as 500 steps during training. The output of the RNN in the statistically transient range ($0\sim50$steps) is not used for generation during training and testing. The time step, $t$, is a random integer larger than 50, which is used for training. The batch size was set to 6, with 2 at each Reynolds number. The training of the RNN-GAN model was carried out for 300 000 iterations.

\begin{figure}
  \centerline{
  \includegraphics[width=0.88\columnwidth]{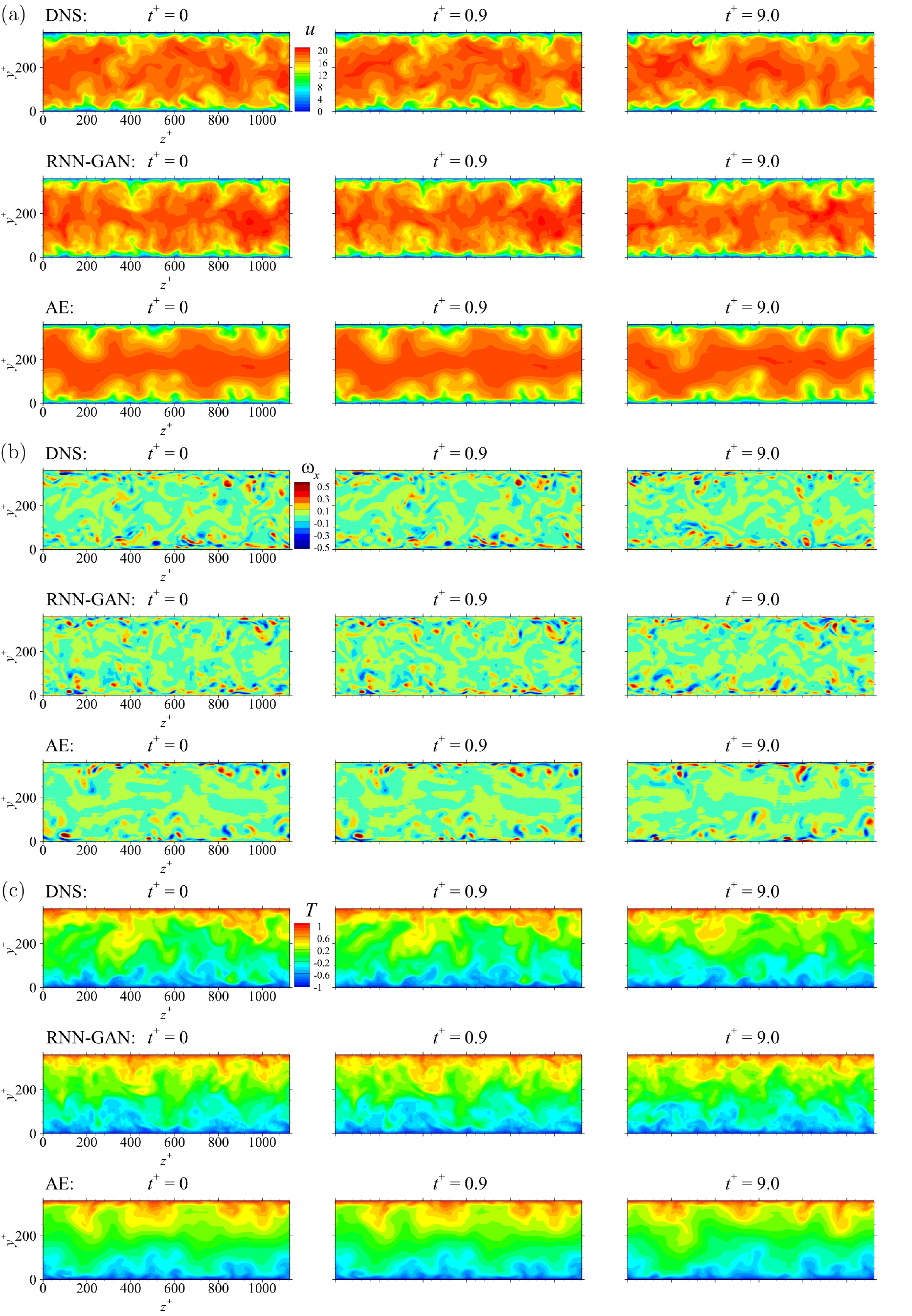}
  }
  \caption{\label{fig:9}{Time-dependently generated flow fields by RNN-GAN at $Re_\tau = 180$. $t^+=0,0.9,$  and $9.0$ corresponding to time steps $t=10000, 10001,$ and $10010$, respectively. (a) Streamwise velocity. (b) Streamwise vorticity normalized by $Re_\tau$. (c) Temperature distribution.}}
\end{figure}

\begin{figure}
	\centerline{
		\includegraphics[width=0.8\columnwidth]{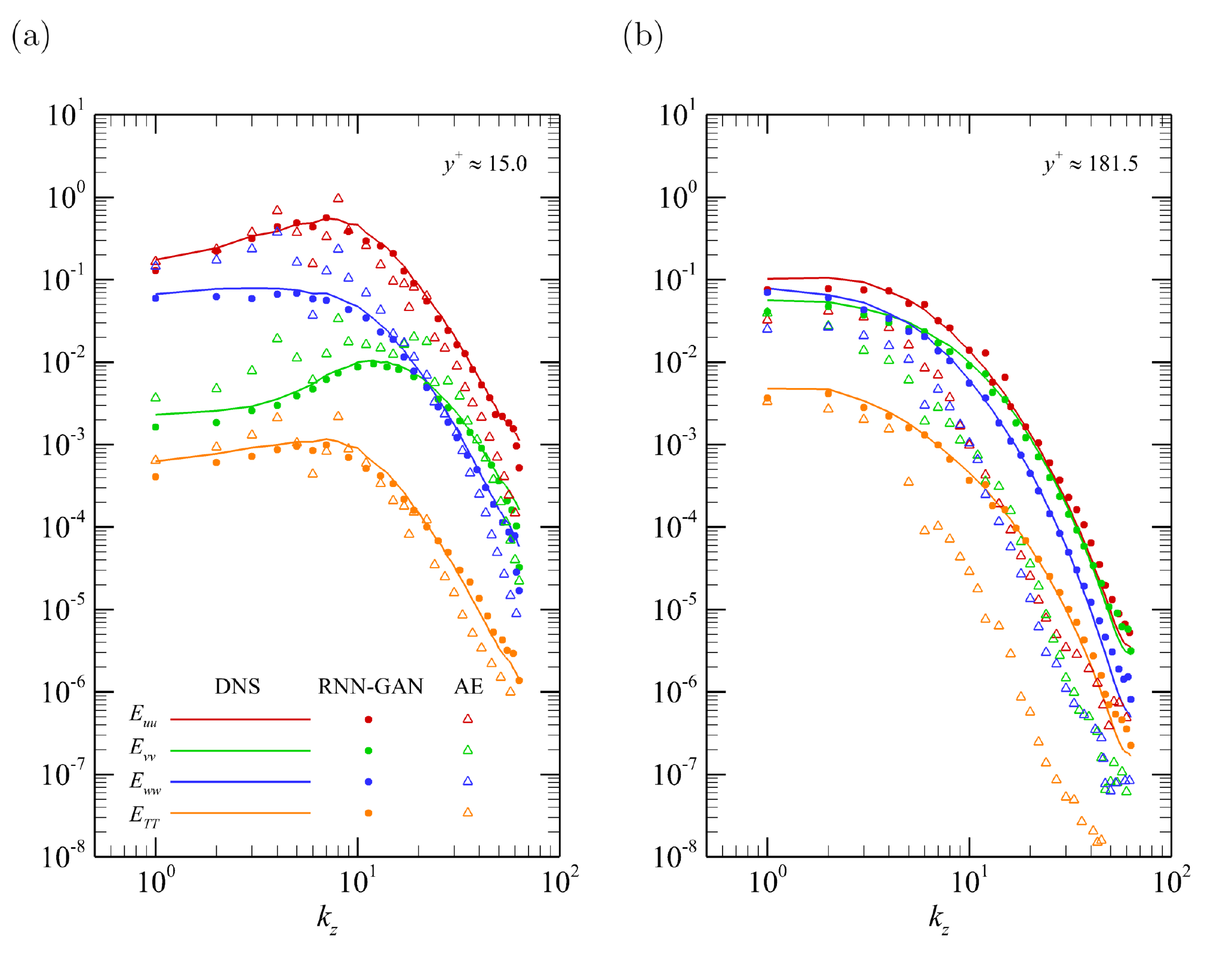}
	}
	\caption{\label{fig:10}{One-dimensional energy spectra results of RNN-GAN at $Re_\tau = 180$. (a)  near the wall, and (b) at the channel center.}}
\end{figure}

\begin{figure}
  \centerline{
  \includegraphics[width=1.05\columnwidth]{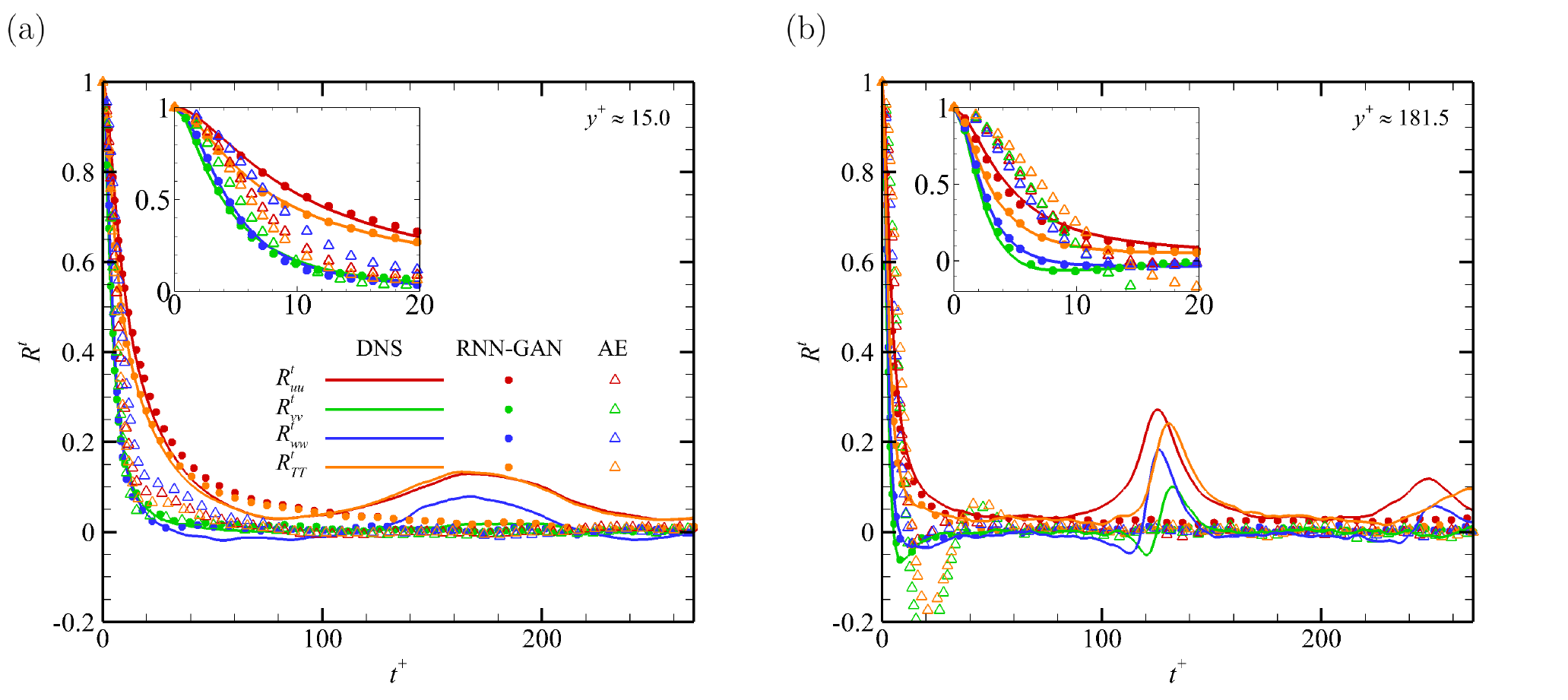}
  }
  \caption{\label{fig:11}{Time correlations of generated flow by RNN-GAN at $Re_\tau = 180$. (a) near the wall ($y^+\approx15$), and (b) at the channel center. Each inset figure shows the magnified view at early time.}}
\end{figure}

Although the training process is slightly complicated, after training, only $G$ and $RNN$ are used for the time-varying flow generation, and it can be simply described as follows:
\begin{equation}
\tilde{X}_{t} = G(Re_\tau, \tilde{z}_{t}) 
\end{equation}
where $\tilde{z}_t =RNN(r_{t},h_{t-1},c_{t-1})$ and $r_{t}$ is the random noise vector from a normal distribution, and the $h_0$ and $c_0$ components are all zero. The flow fields generated by the trained RNN-GAN after 10 000 time steps, approximately 20 times longer than the trained time length ($t_{max}$), are shown in Fig. \ref{fig:9}. Here, 10 000 time steps correspond to 9 000 in wall time units. For the purpose of comparison, the training of AE \cite{Fukami2019}, a supervised learning method, was carried out. The encoder and decoder in the AE consist of a similar number of trainable parameters in the discriminator ($D$) and generator ($G$) of our GAN, respectively. A major difference with regard to the training data is that the AE trained only data at $Re_\tau = 180$ because the AE trained with data at three Reynolds numbers yielded worse results. Detailed information of the AE structure is presented in \ref{app:C}. The fields generated by the AE after 10 000 time steps and those of the DNS are  shown in Fig. \ref{fig:9}. Even after a long time, the RNN-GAN could produce a statistically stationary flow with small-scale turbulent structures near the wall similar to those of the DNS, while the AE generated poorly resolved turbulent flow fields. The spanwise energy spectrums near the wall and the center, which were time-averaged for $50 \le t \le 20050$ excluding the transient period ($1 \le t < 50$), are presented in Fig. \ref{fig:10}. A single time step corresponds to 0.9 in wall time units; therefore, the averaging period is sufficiently long  (as much as 18 000 in wall time units). There is only a slight difference between the spectra of the RNN-GAN and the DNS in the high wave numbers, while the spectrum of the AE shows a large deviation from that of the DNS. The RNN-GAN resolved the spatial correlations so well that the turbulence introduced into a main simulation would not be dissipated quickly.

The temporal correlations of the velocity components and temperature of the field generated by the RNN-GAN are very favorably compared with those of the DNS in Fig. \ref{fig:11}, while those of AE are in poor agreement with the DNS. The different behaviors of the correlations near the wall and the center are correctly captured by the RNN-GAN. These results are in contrast to those of AE, which shows a very sensitive change \cite{Fukami2019} in the time correlations according to the hyperparameters. The spurious periodicity problem that the temporal correlation rebounds after a certain time occurs in the DNS due to the periodic boundary condition, and the insufficient domain size in the streamwise direction \cite{Quadrio2003} disappears properly in our RNN-GAN model similar to that in the AE. It indicates that in a numerical simulation for collecting training data, it is not necessary to expand the domain size considerably in the streamwise direction. 

\begin{figure}
	\centerline{
		\includegraphics[width=0.9\columnwidth]{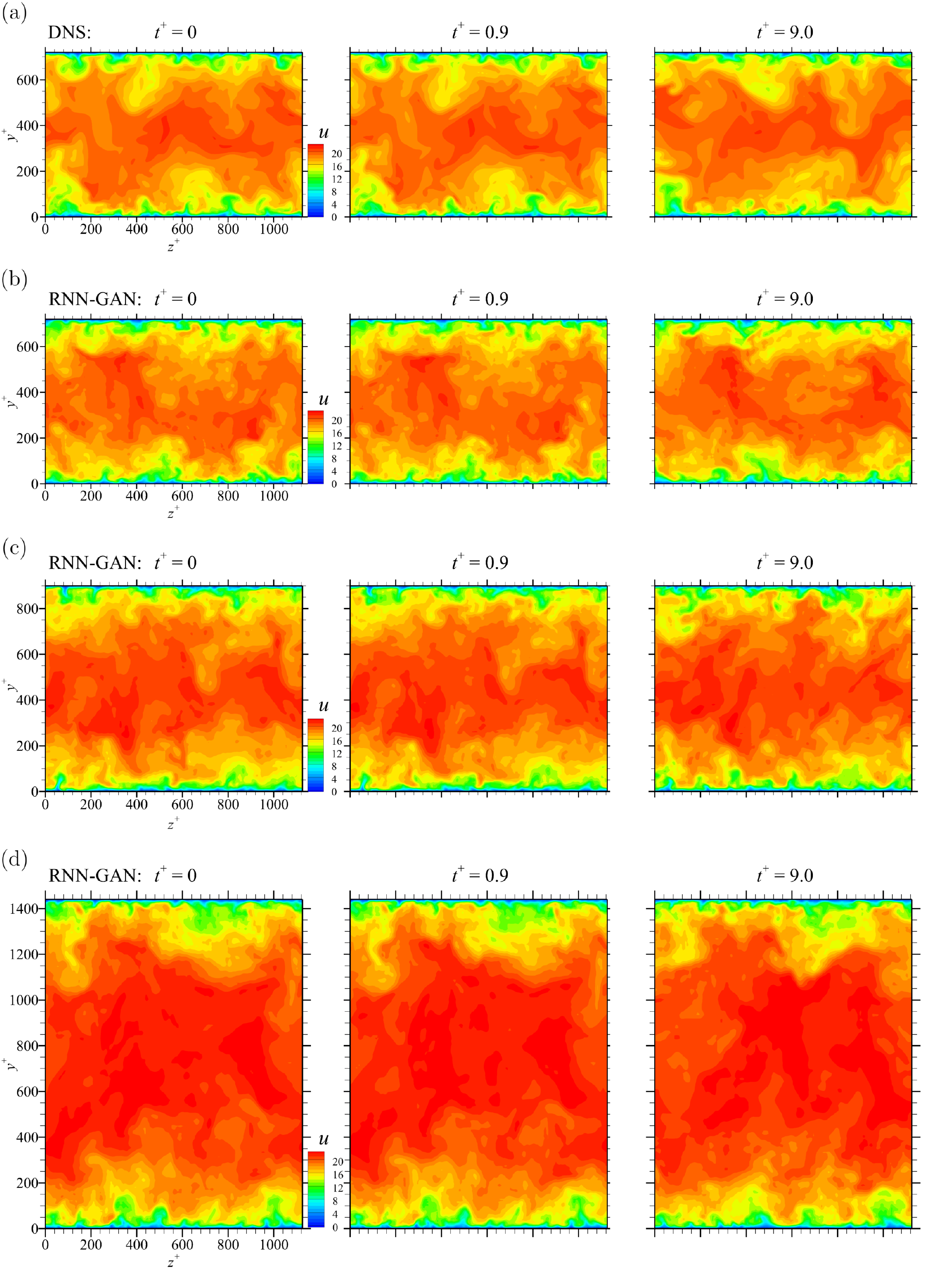}
	}
	\caption{\label{fig:12}{Time-dependently generated streamwise velocity fields by RNN-GAN at various Reynolds numbers. (a,b) are at $Re_\tau=360$, and (c,d) are at $Re_\tau=450,720$, respectively. $t^+=0,0.9,$  and $9.0$ correspond to time steps $t=10000, 10001,$ and $10010$, respectively.}}
\end{figure}

\begin{figure}
	\centerline{
		\includegraphics[width=1.0\columnwidth]{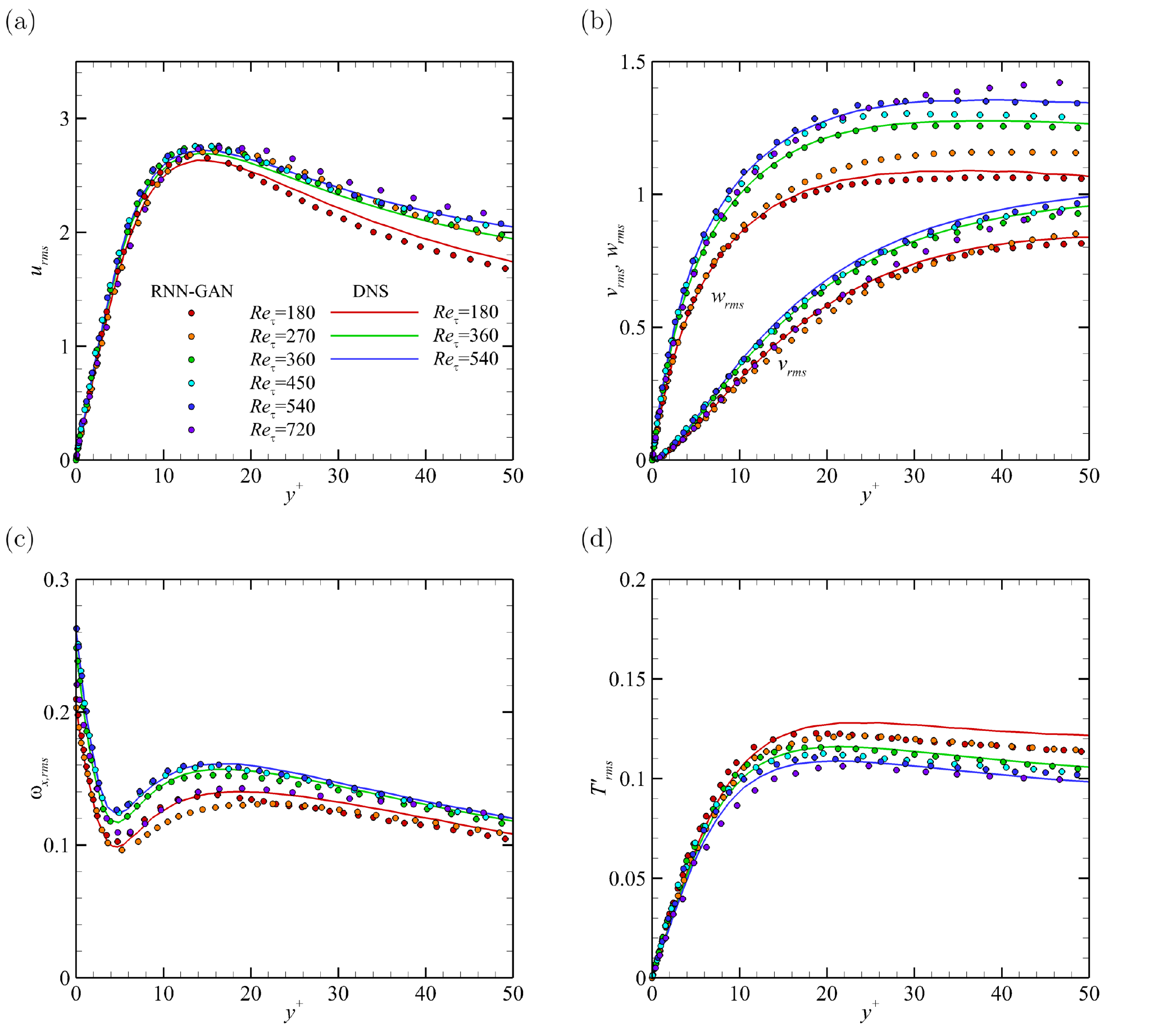}
	}
	\caption{\label{fig:13}{The rms profiles of (a) streamwise velocity, (b) vertical and spanwise velocity, (c) vorticity, and (d) temperature generated by RNN-GAN at various Reynolds numbers.}} 
\end{figure}

In addition to the successful generation of time-varying turbulence fields, our RNN-GAN shows remarkable functions such as flow generation at various Reynolds numbers, stochastic generation of flows, and handling of the variable domain size. First, the RNN-GAN could generate time-dependent flow fields at various $Re_\tau$, as shown in Fig. \ref{fig:12}. In addition to the trained Reynolds number ($Re_\tau=360$), the flow at the other numbers ($Re_\tau=450,720$) could be produced with realistic structures in the near-wall region. Quantitatively, the rms profiles of velocities, vorticity, and temperature generated by the RNN-GAN are compared to those of the DNS in Fig. \ref{fig:13}. Similar to the GAN (Fig. \ref{fig:7}), the RNN-GAN  captures the Reynolds number effect in wall coordinates quite well, although the statistics of the generated fields are not perfectly matched to those of the DNS. Also, the streamwise vorticity, an important variable in the $y-z$ plane of channel flow, of the RNN-GAN is statistically in good agreement with that of the DNS (Fig. \ref{fig:13}(c)). Because the eddies, which are essential in representing the spatial correlation of turbulence, are well expressed, our model could be a good synthetic inflow generator. Regarding the temperature rms profiles that decrease as $Re_\tau$ increases, the RNN-GAN also show similar tendency and accuracy compared with the DNS.

\begin{figure}
	\centerline{
		\includegraphics[width=1.0\columnwidth]{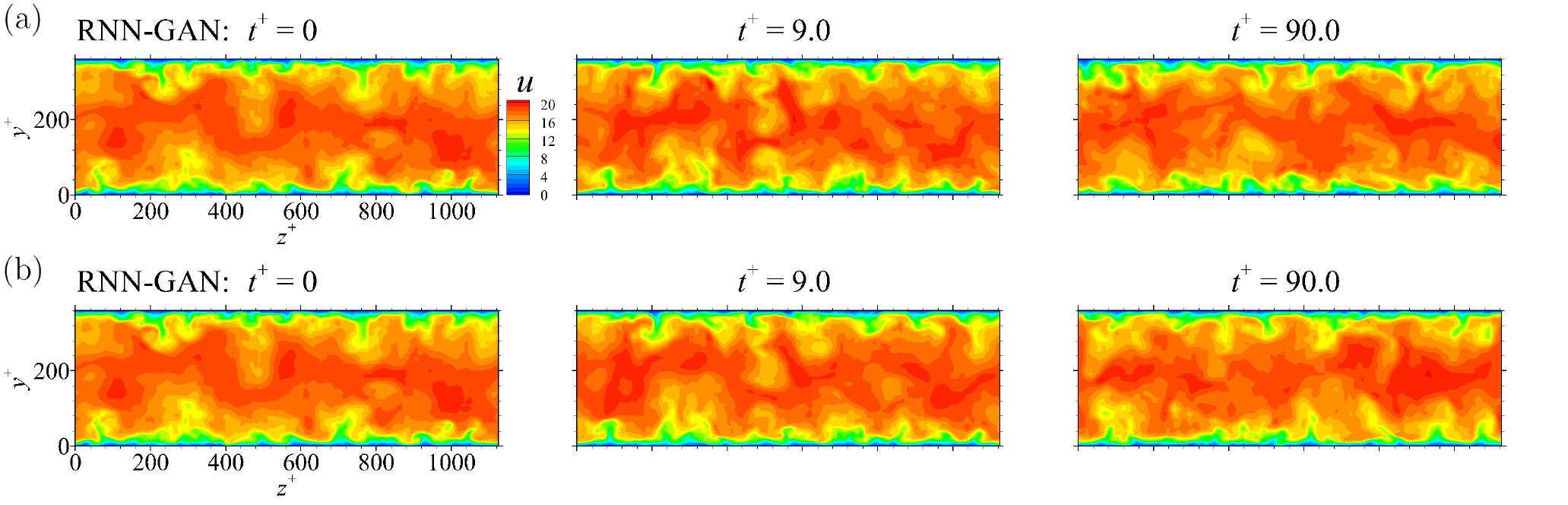}
	}
	\caption{\label{fig:14}{Two flow fields generated by the RNN-GAN with different $\vec{r}_t$ at $Re_\tau=180$.}} 
\end{figure}

Furthermore, in the RNN-GAN, the generated flow could be stochastically varied from the random-noise input of RNN ($\vec{r}_t$). With the same $\vec{r}_t$ where $1\le t\le10000$, the same fields ($\tilde{X}_{10 000}$) are generated. However, the flow becomes different due to the different $\vec{r}_t$, where $10001\le t$ as shown in Fig. \ref{fig:14}. The two flows are partially different when 10 steps have passed. When 100 steps have elapsed, however, it is difficult to find the correlation between them because of the accumulated difference. This indicates that the RNN-GAN satisfies the stochastically varying conditions of inlet flow fields.

\begin{figure}
	\centerline{
		\includegraphics[width=0.675\columnwidth]{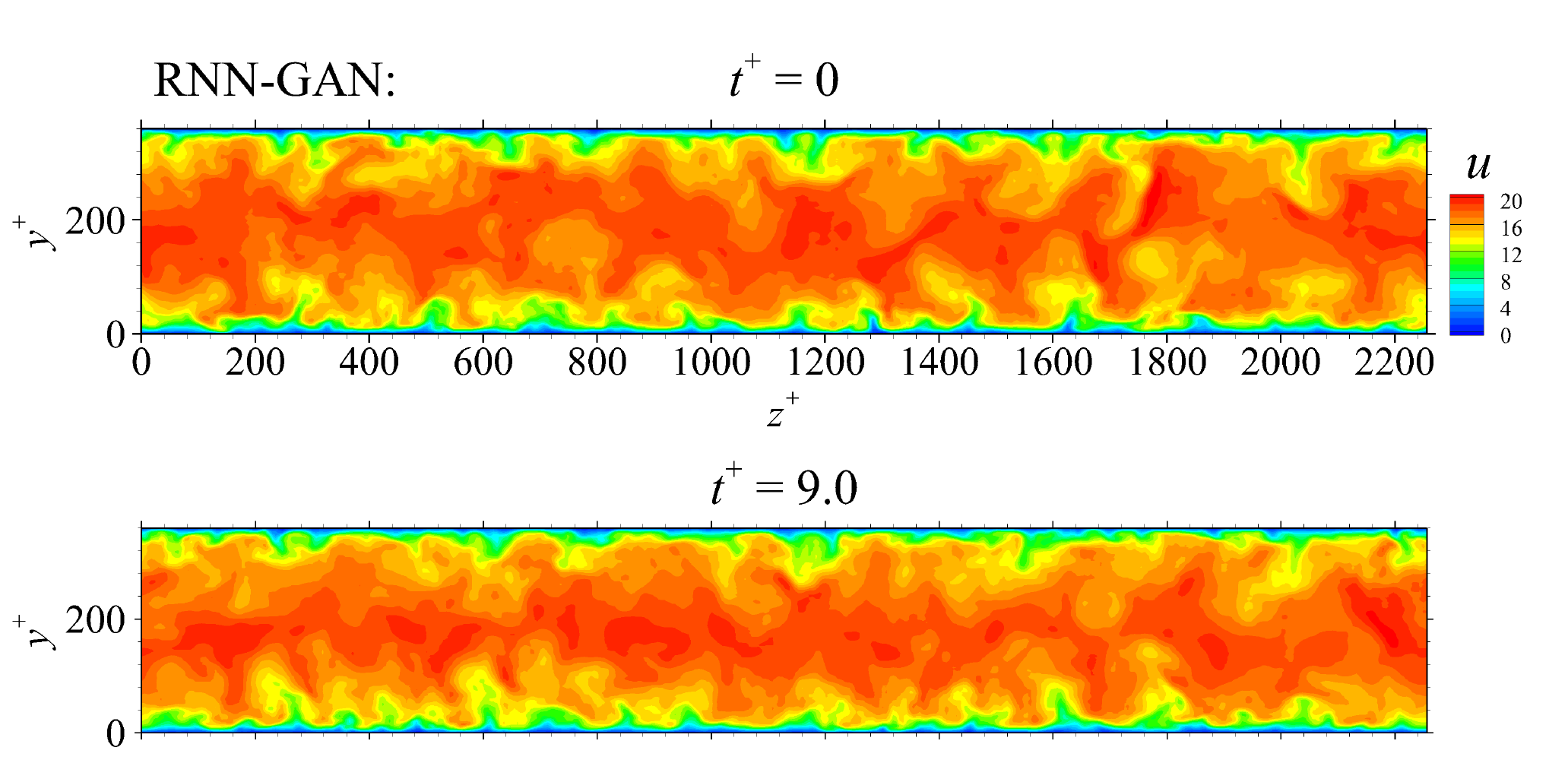}
	}
	\caption{\label{fig:15}{Extended fields in the spanwise direction at $Re_\tau=180$.}} 
\end{figure}

Regarding the inlet domain size, the RNN-GAN has learned the fields with a fixed spanwise length so far. The main simulation, however, might require an inflow condition with a different domain size. The RAN-GAN could easily generate the flow with a different size using the generated one. Fig. \ref{fig:15} shows the generated field with twice larger spanwise length than the trained one. This could be achieved by the combination of two independent vectors in the latent space. A $\tilde{z}_t$ produced by the RNN becomes an image in the first layer of $G$. Then, we attached the image with another independent image in the spanwise direction. As a result of the convolution and upsampling operations in the $G$, a twice larger field could be generated. Of course, large-scale motions, which were not captured by our DNSs, cannot be represented. Although numerical artifacts could occur in the adjacent parts of the two images, we could not observe them. Therefore, we confirmed that the RNN-GAN can generate inflow with various domain sizes.

Finally, the time interval of the main simulation might be different from that of the trained RNN-GAN. Therefore, certain interpolation methods are required to generate the fields between two consecutive generated fields. We recommend that it would be better to generate the flow field through the interpolated input vector between the two input vectors used to generate the two flow fields, rather than to interpolate between the two flow fields generated by the network. For example, we recommend to use $G( \alpha \tilde{z}_{1}+(1-\alpha) \tilde{z}_{2})$ instead of $\alpha G(\tilde{z}_{1})+(1-\alpha) G(\tilde{z}_{2})$. Sometimes, an interpolation between the generated flow fields, the interval of which is too long, can weaken the turbulent structures.

In summary, our results indicate that the proposed RNN-GAN model could time-dependently generate statistically stationary turbulent flow, which is available for an inlet boundary condition of channel flow. Also, when data at a different Reynolds number from the trained one are needed for the inlet boundary condition in an inflow-outflow simulation, the trained RNN-GAN model can generate the corresponding data immediately. The model might be able to reflect other simulation parameters such as Prandtl number, and our framework could be easily expanded to an inflow generator of boundary layer flow.

There are some issues still remaining to be discussed. It could be observed that the spatial correlation of the flow fields generated by the RNN-GAN is a little worse than that generated by the GAN in the generation of an instantaneous field. We expect that an increase in the number of training data, fine-tuning of hyperparameters (especially, increase in the batch size), or further development of RNN and GAN would improve the performance of the present inflow generator. Another issue is that the Reynolds number effect is not perfectly reflected on the GAN, and the extrapolation beyond the trained range is not as good as the interpolation within the trained range. To develop a more general model, a better extrapolation scheme would be essential. As a solution to that problem, the use of a loss function based on the well-known physical constraint, such as the law of the wall, might be helpful for learning the effect, although we did not investigate it here. 

Lastly, the computational cost required for the current model is somewhat demanding. Our RNN-GAN consists of approximately 17M (millions) trainable parameters for the purpose of obtaining the best quality of inflow. There are 7M parameters of $G$, 0.9M parameters of $RNN$, 7M parameters of $D$, 2M parameters of $D_{time}$, and 0.2M parameters of $D_{norm}$. A successful training takes about three weeks on a single GPU machine (NVIDIA Titan Xp), and most of the time was spent for the training of the time-varying flow, in which the 3D convolution operation of $D_{time}$ and the recurrent one of $RNN$ were carried out. It can be reduced by decreasing the number of feature maps in the convolution layers, while it might result in a drop in inflow quality. In this trade-off relation, finding an optimal point, which maintains a good quality at a reasonable learning cost, is an important issue that remains to be solved for the successful application of deep learning to turbulent simulations.   

\section{Conclusions}\label{sec:conclusions}

In this study, we presented an RNN-GAN model based on unsupervised learning that can generate an inlet boundary condition required for developing turbulent flow simulation, as one of the synthetic inflow generators. A previously proposed AE model \cite{Fukami2019} based on supervised learning can also be used as an inflow generator, but small-scale variations of generated turbulence are not well represented and the statistics of the generated fields are inaccurate. In addition, the model requires a fully developed DNS field as an initial input field; therefore, a new DNS should be carried out to get the initial field at a different Reynolds number.  Recognizing these problems, we developed an RNN-GAN model that can generate statistically stationary and more accurate flows at various Reynolds numbers without the initial input field.

First, we trained the GAN model to arbitrarily generate instantaneous flow fields in the cross-sectional $(y-z)$ plane of turbulent channel flow based on the WGAN-GP loss. The generator ($G$) and discriminator ($D$) in the GAN are networks composed of the convolution layers, and many techniques proposed by \citet{Karras2017} are applied in our networks. In order to improve the convergence speed of the GAN and the statistical accuracy of the generated flow, the statistical constraints \cite{Wu2019} were added to the original loss function. Here, we used the energy spectrum of generated fields and the training fields. As a result, an instantaneous flow field generated by the trained GAN was found to have similar characteristics to the DNS field, and all the statistical profiles of the GAN were in very good agreement with those of the DNS. The reason that unsupervised learning of turbulence works well might be related to the existence of statistical characteristics underlying the turbulence. Second, we reflected the Reynolds number effect on the GAN. In order to make a component in $\vec{z}$, which is the input of the $G$, have the meaning of the Reynolds number, the $Re$ map full of the corresponding Reynolds number was used as the input of $D$. After learning the data at three Reynolds numbers, the $G$ could produce the flow fields at various Reynolds numbers and reflect the universal effect of the Reynolds number quite well. Also, the observation of generated fields at several Reynolds numbers indicates that the GAN learned the similarity in the turbulence regardless of the Reynolds number. Of course, the $G$ could not generate a statistically perfect field at a Reynolds number beyond the trained one. Although not tested in this study, it is expected that the changes in other non-dimensional simulation parameters and other simulation conditions would be reflected on the GAN.

Eventually, we attempted to extend the model for instantaneous flow fields to the model for producing time-varying flow fields. We assumed that the trained $G$ could make the time-variation of flow through the change in $\vec{z}$, the input of $G$. Here, the RNN was used to make the change in $\vec{z}$. The input of the RNN is the random noise vector ($\vec{r}_{t}$) that reflects the stochastic variation between time-sequential flow fields ($X_{t-1}$ and $X_{t}$), and the output of the RNN is $\tilde{z}_{t}$, which is expected to generate a flow field ($\tilde{X}_{t}$) at a specific time step through $G$. We expected that $G$ could produce temporally successive flow fields through the time advancement of the RNN. Discriminators including $D$, $D_{time}$, and $D_{norm}$ were used for training the $RNN$-$G$ network. $D$ was used for a successful transfer learning of $G$. $D_{time}$ was used to make $RNN$-$G$ represent the time-variation of flow, and received temporally successive flow fields as input. $D_{norm}$ was used to make the output of the RNN have a normal distribution so that the generated flow is statistically stationary for a long time. After training of the constructed RNN-GAN including $G$, $RNN$, $D$, $D_{time}$, and $D_{norm}$, we found that the trained network could generate fully-developed flow fields at various Reynolds numbers for a very long time similar to those of the DNS, and  both time correlations and spanwise energy spectrum of the generated fields were in good agreement with those of the DNS. In addition, the RNN-GAN has some characteristics that the generated flow is stochastically varying, and the domain size of the generated field can be varied.

In summary, our study indicates that the proposed RNN-GAN model can be a successful synthetic inflow generator for producing an inlet boundary condition of turbulent channel flow, which is necessary to carry out a developing flow simulation. This framework, which is based on the unsupervised learning, is likely to be extended to the inflow generation of the boundary layer flow or other simulations. In the present work, we showed that deep learning based on GAN could be a useful tool for turbulence simulations.

\vspace{0.3in}
\noindent
{\bf {Acknowledgments}}

This work was supported by the National Research Foundation of Korea (NRF) grant funded by the Korea government (MSIP) (2017R1E1A1A03070282).

\appendix

\section{Effect of the statistical constraint}\label{app:A}

\begin{figure}
	\centerline{
		\includegraphics[width=1.0\columnwidth]{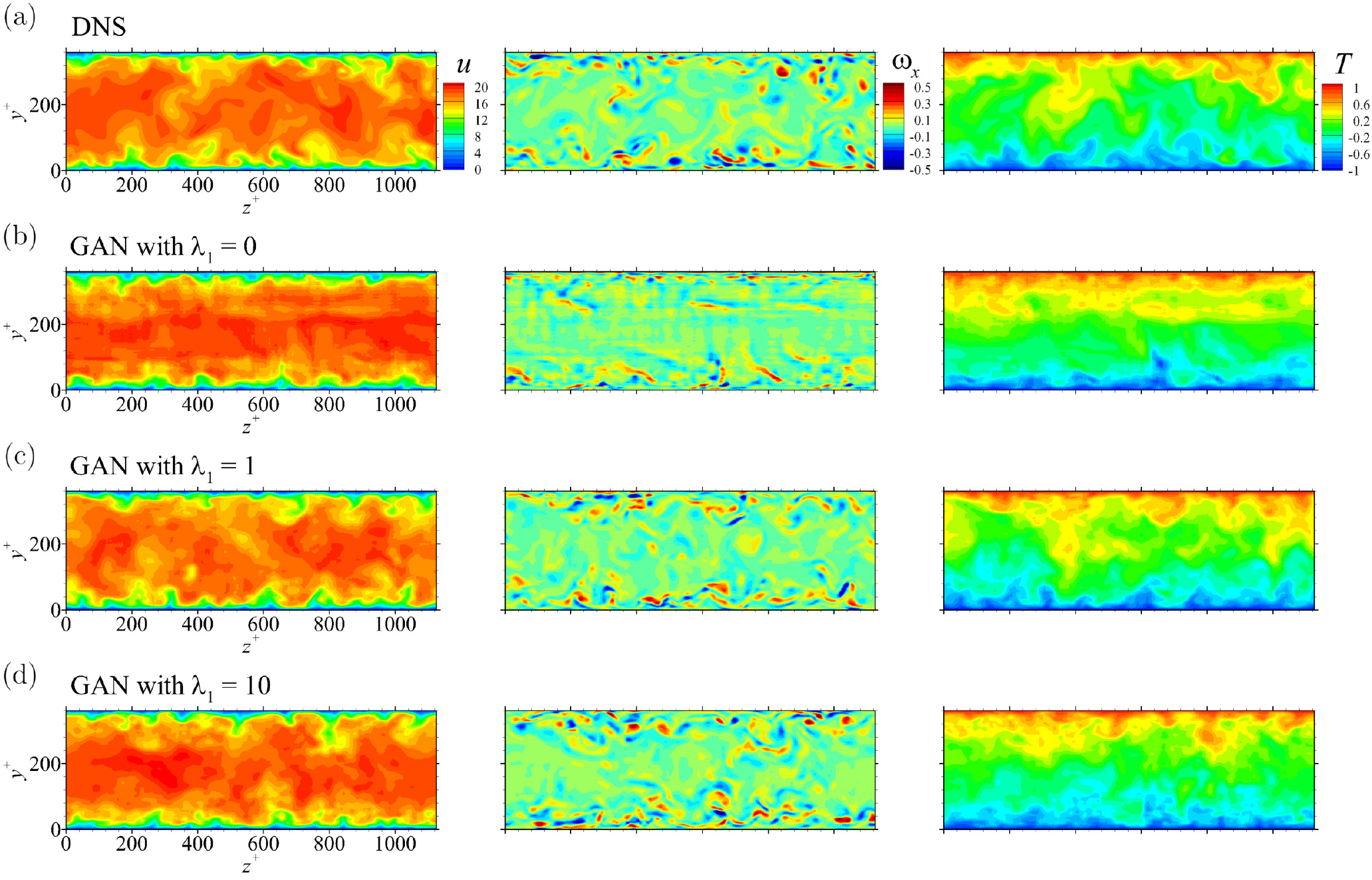}
	}
	\caption{\label{fig:17}{Generated fields by (a) DNS at $Re_\tau=180$, (b) GAN without the statistical constraint ($\lambda_{1}=0$), (c) GAN with a weaker statistical constraint ($\lambda_{1}=1$), and (d) GAN with a stronger statistical constraint ($\lambda_{1}=10$). All trainings were carried out for short 30 000 iterations.}}
\end{figure}

Our proposed model was trained with the statistical constraint \cite{Wu2019}. In particular, in our work, the energy spectra calculated in the wave space was used as the constraint. This effectively worked not only to improve the quality of the generated fields but also to speed up the learning process. The effect of constraint strength, $\lambda_{1}$, at 30 000 training iterations is shown in Fig. \ref{fig:17}. The GANs with the constraint ($\lambda_{1}=1$ and $10$) were trained much faster and produced a better quality field than the GAN without it ($\lambda_{1}=0$). For the training of RNN-GAN, the effect of the constraint became very significant (not shown here). Also, the greater strength ($\lambda_{1}=10$) prevented the sudden deterioration in the transfer learning of RNN-GAN better than the weaker one ($\lambda_{1}=1$), although the difference was not large. According to the type of statistics, there is an optimal strength value so that a small tuning process is needed. Nevertheless, the statistical constraint is recommended for the application of GAN in the turbulence research because of the great advantage, speed, and quality.

\section{Effect of data augmentation}\label{app:B}

\begin{figure}
	\centerline{
		\includegraphics[width=1.0\columnwidth]{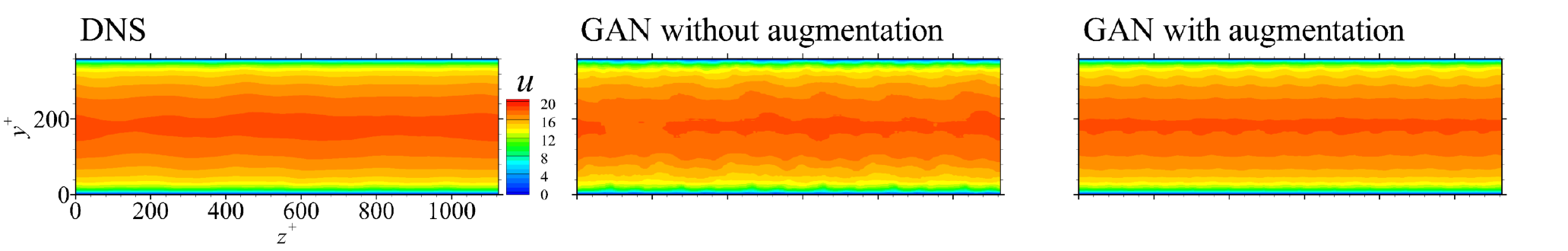}
	}
	\caption{\label{fig:16}{Pointwise-averaged streamwise velocity of DNS, GAN without augmentation schemes, and GAN with mirror- and spectral augmentation schemes at $Re_\tau=180$. All the trainings were carried out for 100 000 iterations.}}
\end{figure}

In the present study, we used data augmentation schemes including mirror- and spectral augmentation, because in machine learning, the amount of training data is the most critical factor affecting the training results. In particular, mirror-augmentation was expected to reflect the statistically symmetric characteristic of channel flow on the GAN, and the spectral one was expected to reflect the statistically homogeneous property on the GAN. We trained the two GANs, one with augmentation and one without it, to observe the effect of  augmentation on the training results. After training for 100 000 iterations, the streamwise mean velocity of the fields generated by the GANs are as shown in Fig. \ref{fig:16}. The statistics were ensemble-averaged and pixelwise-averaged using 10 000 randomly generated fields. The statistic of DNS was time-averaged using the time-series of 20 000 training fields without spanwise average and any augmentation. The results show the difference in fluctuations in the spanwise direction, indicating that the GAN with augmentation represents the statistically symmetric and homogeneous characteristics much better than the GAN without it. Also, the augmentation made the learning of GAN stable, especially at the early training steps. In addition to our approaches, physically reasonable data augmentation might improve the training results.

\section{Inflow generation based on the supervised learning}\label{app:C}

We conducted training of a supervised learning model \cite{Fukami2019} for the purpose of comparison with our approach. \citet{Fukami2019} reported that the inflow generation using the AE model, which predicts the next step field based on the previous step one, is possible. The AE can generate statistically stationary flow fields of channel flow through recursive prediction. However, this model has considerable problems with respect to the spatial and temporal correlations of flow. Also, it requires a fully-developed DNS field as the initial input; hence, the model has a limited utility. 

This architecture consists of an encoder that takes the the previous step field as input and compresses it into a low dimension, a multilayer perceptron (MLP) that is composed of fully-connected layers in the low dimension, and a decoder that reconstructs the next step field from the low dimension. The encoder and decoder of the trained model are composed of convolution layers to efficiently handle the spatial correlations of the flow field, and the size of the dimension changes through downsampling and upsampling in the encoder and the decoder, respectively. This network can be described as follows:
\begin{equation}
{X_{t+1} = AE(X_{t}) = DEC(MLP(ENC(X_{t})))}
\end{equation}
where $DEC$ and $ENC$ are the decoder and encoder networks, respectively. In the present study, $X_{t}=(u,v,w,T)$. The DNS data is used as the initial input field ($X_{0}$). The training of the constructed network is performed only when $t = 0$, and the loss function can be defined as the mean square error ($\textnormal{MSE}$) between $X_{1}^{AE}$ and $X_{1}^{DNS}$,

\begin{equation}
{\textnormal{MSE} = {1 \over N}\sum_{i=1}^{N}({X_{1}^{i,AE}-X_{1}^{i,DNS}})^2}
\end{equation}
where $N$ is the number of datasets per training iteration, and $X_{1}^{i,AE}$ and $X_{1}^{i,DNS}$ are the predicted field and the DNS field, respectively. 

An inflow generator based on supervised learning aims to obtain information when $t \geq 1$ through recursive prediction after training. A classical CNN has a problem that the prediction accuracy for $t > 1$ is not guaranteed and decreases with an increase in $t$, because the input field at $t \geq 1$ is worse than the DNS field. However, in the AE, even if the difference between $X_{t}^{AE}$ and $X_{t}^{DNS}$ is large, the difference between $ENC(X_{t}^{AE})$ and $ENC(X_{t}^{DNS})$ is small so that this problem is prevented. In other words, the compression process of encoding prevents the network from being overfitted to $t = 0$. Therefore, through the recursive prediction using the trained AE, statistically stationary time-varying flow fields can be obtained.

In order to carry out a fair comparison with our proposed model, the training of the AE was carried out with a similar number of trainable parameters in the encoder and decoder of the AE as that in the discriminator and generator of the proposed model, respectively. The discrete convolution using $3\times3$ kernel is applied in the encoder and decoder. Periodic padding is used instead of zero padding in the spanwise directional convolution operation so that the periodic boundary condition is naturally reflected. Also, $2\times2$ average pooling and nearest-neighbor interpolation are used for downsampling and upsampling, respectively. The MLP in the compressed latent space consists of two fully-connected layers. Here, the number of nodes in the hidden layer is related to the compression ratio that has a decisive influence on the generator performance. A large number of nodes makes the generator overfitted to $t=0$, whereas, a small number of nodes makes the generated field inaccurate. The best case among several tests that we carried out is composed of 768 nodes in the hidden layer. It should be noted that 768 is larger than the vector size of the latent space in our proposed model. As a nonlinear function in the AE, the hyperbolic tangent function (tanh) is applied \cite{Fukami2019}. Only the data at $Re_\tau=180$ was used for training, and the time interval of the temporally successive training fields ($X_{0}^{DNS}$ and $X_{1}^{DNS}$) was $0.9$ which $\Delta t^{+}_{AE}$ is expected to be. To minimize $\textnormal{MSE}$, ADAM was used. The batch size ($N$) was set to 8, and training was performed for a total of 300 000 iterations in which the validation error did not decrease anymore. The initial learning rate was set to 0.0001, and it decreased by $1\over10$ per 100 000 iterations. The validation error was monitored during training, and the overfitting, in which the validation error and the training error differ greatly, did not occur.

\begin{figure}
	\centerline{
		\includegraphics[width=1.0\columnwidth]{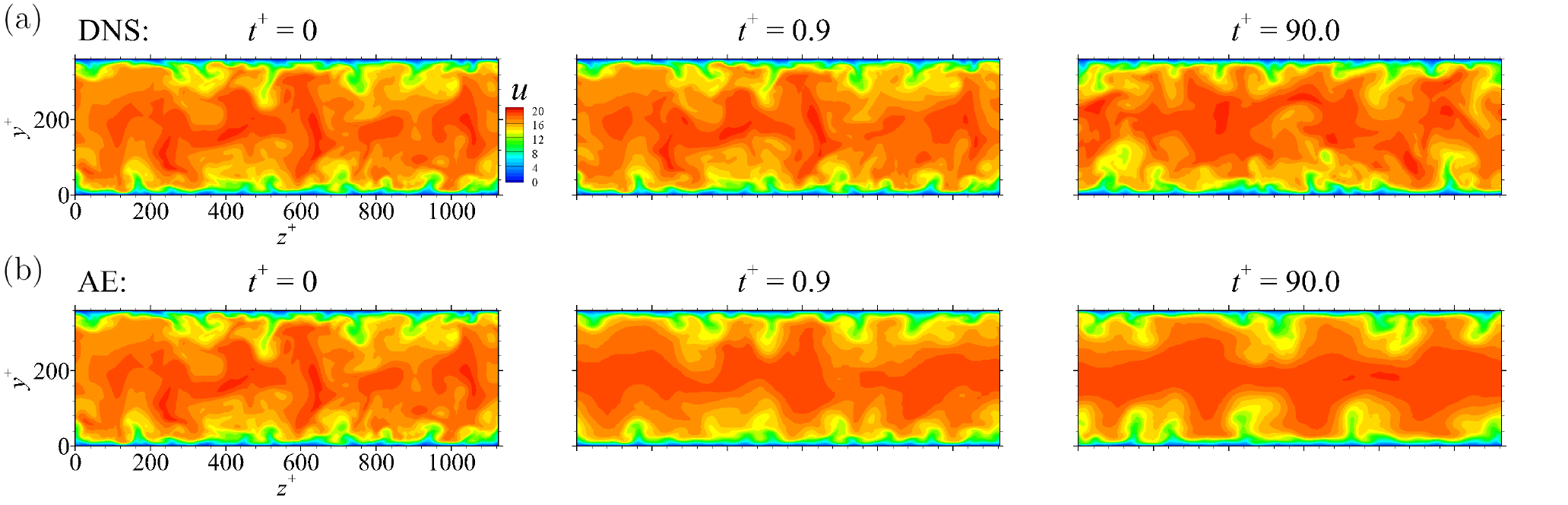}
	}
	\caption{\label{fig:18}{DNS and generated flow fields by the autoencoder(AE) network based on supervised learning. (a) DNS results, and (b) AE results. The initial input field ($t^+=0$) of the autoencoder is the DNS field ($t^+=0$). 0,0.9, and 90.0 in wall time units ($t^+$) correspond to 0, 1, and 100 time steps of AE, respectively.}} 
\end{figure}

After the above training process, flow fields were generated by the trained AE using a new initial field, which is far  from the training fields, as shown in Fig. \ref{fig:18}. Even after a single time step ($t^+ = 0.9$), the generated field was poorly resolved. The relatively large-scale structures are similar to those of the DNS, but small-scale variations are not well represented. Also, after 100 time steps ($t^+ = 90.0$), it is difficult to find similarity between the generated field and the DNS field because the difference from the DNS accumulated with time advancement. The quality of the output field at $t^+ = 90.0$, obtained through recursive prediction to input the generated field again, does not deteriorate significantly compared to the field at $t^+ = 0.9$. 

The time correlations, which were time-averaged over 20000 time steps ($\Delta t^{+} = 18000$), were compared with those of the DNS. At the range from $t^+=0$ to $t^+=20$, they were very different from those of the DNS (Fig. \ref{fig:11}). Furthermore, the training results change sensitively according to the hyperparameters in the MLP \cite{Fukami2019}. A possible reason for this problem is that this AE considers the time-variation only for one time step. In fact, we trained a network that uses the previous 3 step information as input to better express the temporal correlation, but it was rather overfitted in the first time step ($t=0$) and showed worse results than before. Also, other statistics including the mean, root-mean-square (rms), and pointwise correlation, show some errors, compared with our proposed model (not shown here). In summary, the AE based on supervised learning does not accurately represent the spatiotemporal correlations of the generated flow, although recursive prediction is possible for a long time.

In addition, we tested whether the AE can learn and generate data at various Reynolds numbers. However, the trained AE showed much poorer rms profiles than those of the AE that was trained using only a Reynolds number data. Also, for generation at a different Reynolds number from the trained one, this method is inefficient because it requires a fully developed field at the corresponding Reynolds number as the initial input. In the present study, our model, RNN-GAN based on the unsupervised learning, solved the above problems.

\bibliographystyle{model1-num-names}
\bibliography{refs}

\end{document}